\documentclass[aps,prx,twocolumn,superscriptaddress]{revtex4-2}
\usepackage[english]{babel}
\usepackage{amssymb}
\usepackage{amsmath}
\usepackage{mathtools}
\usepackage{txfonts}
\usepackage{mathdots}
\usepackage[normalem]{ulem}
\usepackage{array}
\usepackage{amsfonts}
\usepackage{mathrsfs}
\usepackage{booktabs}
\usepackage{threeparttable}
\usepackage{multirow}
\usepackage[dvips]{graphicx}
\usepackage{epsfig}
\usepackage{graphicx}
\usepackage{subfigure}
\usepackage{threeparttable}
\usepackage{chngpage}
\usepackage{float}
\usepackage{xcolor}
\usepackage{bm}
\usepackage[toc]{appendix}
\usepackage{tabularx}
\usepackage{adjustbox}
\usepackage{comment}
\usepackage{hyperref}
\usepackage{lineno}
\hypersetup{
    unicode=false,     
    pdftoolbar=false,  
    pdfmenubar=true,   
    pdffitwindow=false, 
    pdfstartview={FitH},
    pdftitle={},    
    pdfauthor={Authors},     
    pdfsubject={},   
    pdfcreator={},   
    pdfproducer={}, 
    pdfkeywords={quantum state transfer} {superconducting processor} {quantum state tomography}, 
    pdfnewwindow=true,
    colorlinks=true,
    linkcolor=black,
    citecolor=blue, 
    filecolor=magenta,
    urlcolor=blue
}

\newcommand{\tcr}{\textcolor{red}}

\begin{document}
\nolinenumbers
\title{Enhanced quantum state transfer: Circumventing quantum chaotic behavior}
\author{Liang Xiang}
\thanks{These authors contributed equally}
\author{Jiachen Chen}
\thanks{These authors contributed equally}
\author{Zitian Zhu}
\thanks{These authors contributed equally}
\author{Zixuan Song}
\author{Zehang Bao}
\author{Xuhao Zhu}
\author{Feitong Jin}
\author{Ke Wang}
\author{Shibo Xu}
\author{Yiren Zou}
\author{Hekang Li}
\author{Zhen Wang}
\author{Chao Song}
\affiliation{School of Physics, ZJU-Hangzhou Global Scientific and Technological Innovation Center, and Zhejiang Province Key Laboratory of Quantum Technology and Device, Zhejiang University, Hangzhou, China}
\author{Alexander Yue}
\affiliation{Department of Physics and Astronomy, University of California, Davis, CA 95616, USA}

\author{Justine Partridge}
\affiliation{Department of Physics and Astronomy, University of California, Davis, CA 95616, USA}

\author{Qiujiang Guo}
\email{qguo@zju.edu.cn}
\affiliation{School of Physics, ZJU-Hangzhou Global Scientific and Technological Innovation Center, and Zhejiang Province Key Laboratory of Quantum Technology and Device, Zhejiang University, Hangzhou, China}
\author{Rubem Mondaini}
\email{rmondaini@csrc.ac.cn}
\affiliation{Beijing Computational Science Research Center, Beijing 100193, China}

\author{H. Wang}
\affiliation{School of Physics, ZJU-Hangzhou Global Scientific and Technological Innovation Center, and Zhejiang Province Key Laboratory of Quantum Technology and Device, Zhejiang University, Hangzhou, China}

\author{Richard T.~Scalettar}
\email{scalettar@physics.ucdavis.edu}
\affiliation{Department of Physics and Astronomy, University of California, Davis, CA 95616, USA}

\begin{abstract}
The ability to realize high-fidelity quantum communication is one of the many facets required to build generic quantum computing devices. In addition to quantum processing~\cite{DiVincenzo2000, Ladd2010}, sensing~\cite{Degen2017}, and storage~\cite{Terhal2015, Brown2016}, transferring the resulting quantum states demands a careful design that finds no parallel in classical communication. Existing experimental demonstrations of quantum information transfer in solid-state quantum systems are largely confined to small chains with few qubits~\cite{Chapman2016, Li2018, Kandel2021, Kandel2019, Sigillito2019, Qiao2020, Mills2019, Yoneda2021, Nakajima2018, vanDiepen2021, Kurpiers2018, Zhong2021, Dumur2021}, often relying upon non-generic schemes. Here, by using a large-scale superconducting quantum circuit featuring thirty-six tunable qubits, accompanied by general optimization procedures deeply rooted in overcoming quantum chaotic behavior, we demonstrate a scalable protocol for transferring few-particle quantum states in a two-dimensional quantum network. These include single-qubit excitation and also two-qubit entangled states, and two excitations for which many-body effects are present. Our approach, combined with the quantum circuit's versatility, paves the way to short-distance quantum communication for connecting distributed quantum processors or registers~\cite{Bose2007review}, even if hampered by inherent imperfections in actual quantum devices.
\end{abstract}

\maketitle

\section*{Introduction}
Among the many desired features of future large-scale quantum computation, the manipulation and transmission of quantum states without destroying their fragile coherence stand out as of primal importance. Originally, the transport of quantum information has been theoretically proposed~\cite{Cirac1997} and experimentally demonstrated~\cite{Matsukevich2004, Bao2012} by using entangled photons to mediate the information transfer between atom clouds over long distances, allowing quantum teleportation of states~\cite{Bouwmeester1997, Ren2017} and the implementation of quantum key-distribution~\cite{Liao2017, Nadlinger2022}, a fundamental step towards the realization of long-distance quantum secure communication~\cite{Portmann2022}. Recently, the growing system sizes of quantum computing platforms~\cite{Ebadi2021, zhang2017nature, Arute2019, Wu2021, Xu2023cpl} make it of paramount importance to realize quantum communication between different parts of a single device (or short-range quantum networks~\cite{LaRacuente2022}), particularly for solid-state architectures with local interactions~\cite{xue2022nature, Arute2019}.

Considering short-distance quantum communication in solid-state devices, implementations in small chains of silicon-based quantum dots have primarily led the way~\cite{Nakajima2018, Kandel2019, Sigillito2019, Mills2019, Qiao2020, Yoneda2021, Kandel2021, vanDiepen2021}. Specific conditions for their realizations include adiabatic tuning of qubit couplings~\cite{Kandel2021}, successive application of SWAP gates~\cite{Kandel2019, Sigillito2019, Qiao2020}, or via shuttling charges either with engineered gate pulses~\cite{Mills2019, Yoneda2021}, assisted by noise~\cite{Nakajima2018} or utilizing Coulomb repulsion~\cite{vanDiepen2021}. Transmission of quantum states, entangled or not, between far apart superconducting qubit nodes can be accomplished via a superconducting coaxial cable~\cite{Kurpiers2018,Zhong2021}, or via surface acoustic wave phonons~\cite{Dumur2021}. 

\begin{figure*}
\includegraphics[width=2\columnwidth]{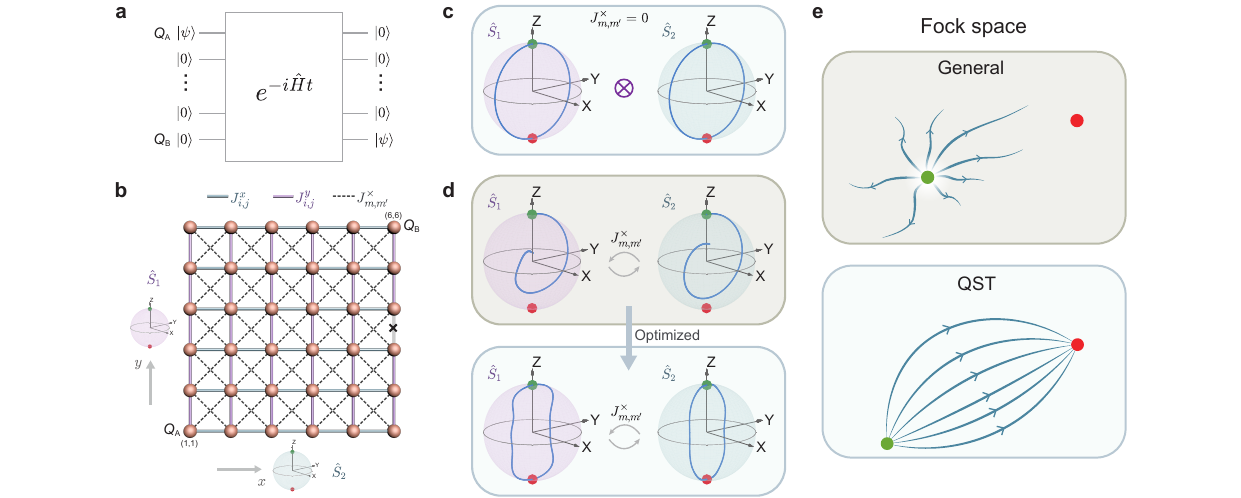}
    \caption{
    {\bf Schematic representation of quantum state transfer.} 
    {\bf a}, The QST process for a single qubit excitation is achieved by finding a suitable Hamiltonian $\hat H$ which at time $t=t_{\rm QST}$ results in the transfer of the state $|\psi\rangle$ initially encoded in qubit $Q_A$ to qubit $Q_B$.
    {\bf b}, Large-spin representation of a QST in a $D=2$ network. Without cross-couplings or defects, the transfer process of a quantum state $|\psi\rangle$ encoded in qubit $Q_A$ to its opposite-symmetric qubit $Q_B$ can be regarded as the independent precession of the spin of two fictitious particles, each mapping a direction of the qubit network; here $N_1=N_2=6$. A perfect QST emerges at $tJ=\pi/2$ when they precess at the same rate $2J$. In the circuit network, nearest-neighbor qubit couplings along the $x$($y$) directions are denoted by  $J_{i,j}^x$($J_{i,j}^y$), whereas $J^{\times}_{m,m^\prime}$ gives the amplitude of the naturally occurring parasitic intraplaquette next-nearest neighbor couplings. The gray bond with cross marker depicts the defect, a malfunctioning coupler in our device.
    {\bf c}, The trajectory $\{\langle\hat S_{i,x}(t)\rangle,\langle\hat S_{i,y}(t)\rangle,\langle\hat S_{i,z}(t)\rangle\}$ in the enlarged Bloch sphere of the two mapped spins, $i = 1, 2$, when the nearest-neighbor couplings of the original network are parametrically selected as $J_{n,n+1}^{x\to 1,y\to 2} = J\sqrt{n (6-n)}$, without cross-couplings ($J^{\times}_{m,m^\prime}=0$) or defect.
    {\bf d},  $J^\times_{m,m^\prime}\neq0$ and defect disturb the perfect precession, breaking the standard protocol~\cite{Christandl2004}, and the desired QST fails. Optimizing couplings $J_{n,n+1}^{1,2}$ compensates for the effects of imperfections, allowing the `wiggled' evolution to achieve QST within desired time scales. {\bf e}, Cartoon contrasting the general picture for the evolution in Fock space of an initial state (green dot) under generic or QST-optimized Hamiltonians. General dynamics tend to be ergodic and quickly diffuse the initial information in the Fock space, while the QST dynamics manifest nonergodic behavior, re-converging to the final target state (red dot) at later times.} 
    \label{fig:1}
\end{figure*}

Heading towards a full quantum digital scheme,  while the sequential application of SWAP gates is the standard candidate for a platform-agnostic transmission of a quantum state, the accumulation of minor two-qubit gate errors, accompanied by associated timing gate errors, can ultimately hinder an efficient quantum state transfer (QST)~\cite{Foulk2022}. An alternative approach, which avoids complex dynamical control of inter-qubit operations, is to use pre-engineered couplings that, in quantum circuits governed by a static Hamiltonian, achieve high-fidelity transfer of quantum information~\cite{Bose2003}---see Fig.~\ref{fig:1}{\bf a}.

\begin{figure*}
\includegraphics[width=2.0\columnwidth]{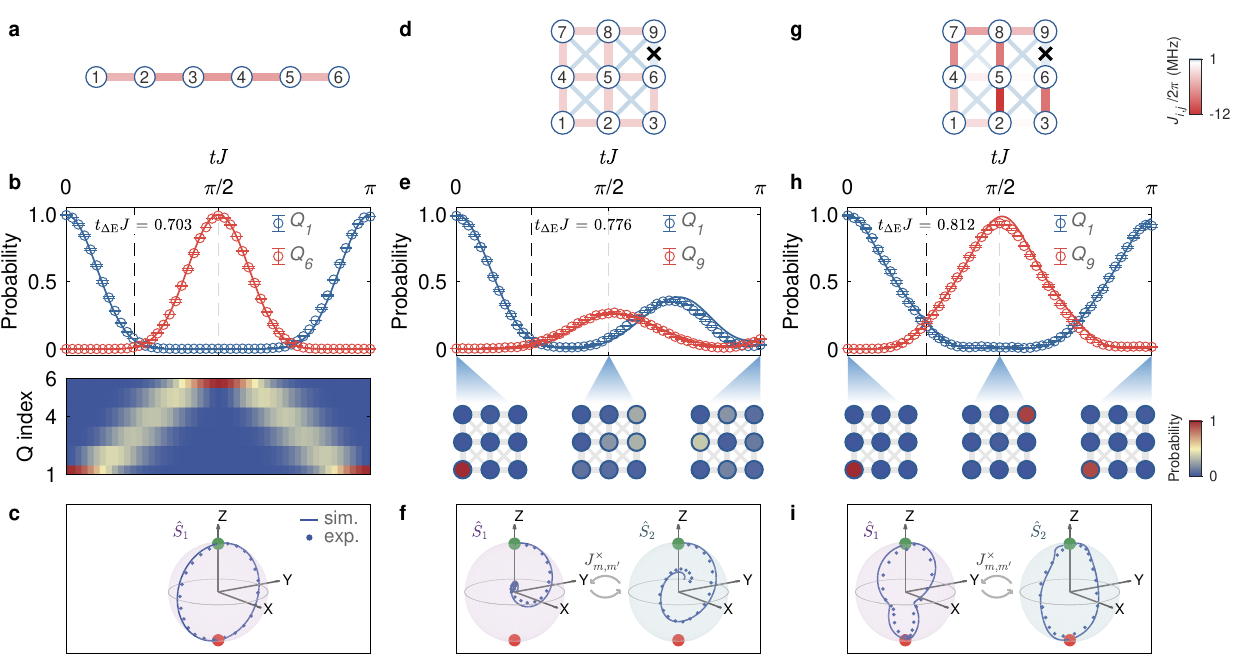}
    \caption{
    {\bf   Single-excitation quantum state transfer in 1D ($1\times6$) and small 2D ($3\times3$) systems.} 
    {\bf a},  Schematic representation of 1D $1\times6$ qubits with the NN couplings parameterized by the standard protocol $J_{n,n+1} = -J \sqrt{n(6-n)}$ with $J/2\pi=2$ MHz~\cite{Christandl2004} . {\bf b}, The corresponding experimental dynamics of $Q_1$ and $Q_6$ excited-state populations (top), whose time-evolution heatmap for all 6 qubits is shown at the bottom. {\bf c}, The associated trajectory of $\langle \hat S_{1,\alpha}\rangle$ ($\alpha=x,y,z$) in the large-spin representation; markers (line) give the experimental (simulated) results. {\bf d}, $3\times3$ qubits with uniform NN couplings, $J^{(1,2)}_{n,n+1}/2\pi \approx -2\sqrt{2}$~MHz  \cite{Christandl2004}, except for the defective coupler, showing the excited population dynamics of qubits $Q_1$ and $Q_9$ [{\bf e, top}] and snapshots for all qubits at representative times [{\bf e, bottom}]; the corresponding spin trajectories, $\langle \hat S_{1,\alpha}\rangle$ and $\langle \hat S_{2,\alpha}\rangle$, describing the lack of perfect precession in the presence of cross-coupling terms  $J^\times_{m,m^\prime}$ and the defective coupler, is shown in {\bf f}. {\bf g}, {\bf h}, and {\bf i} show the same for the case in which we optimize the couplings in the $3\times3$ qubit network to achieve a good QST -- despite the wiggled evolution of the spin-trajectories, the synchronized precession is recovered, and the fidelity of the QST is dramatically improved. The cross marker in the bond connecting qubits $Q_6$ and  $Q_9$ denotes the defective coupler with a fixed value of about $2\pi\times0.3$ MHz. Error bars here come from the standard deviation of five experimental repetitions.  $t_{\Delta E}$ is a minimum time, set by `quantum speed limit' arguments, for the generation of a final state localized on a different site, and hence orthogonal to the initial state --  see Eq.~\eqref{eq:t_qsl}. See Supplementary Section \tcr{2} for the experimentally measured coupling values in \textbf{a}, \textbf{d}, and \textbf{g}.}
    \label{fig:2}
\end{figure*}

\begin{figure*}
\includegraphics[width=2.0\columnwidth]{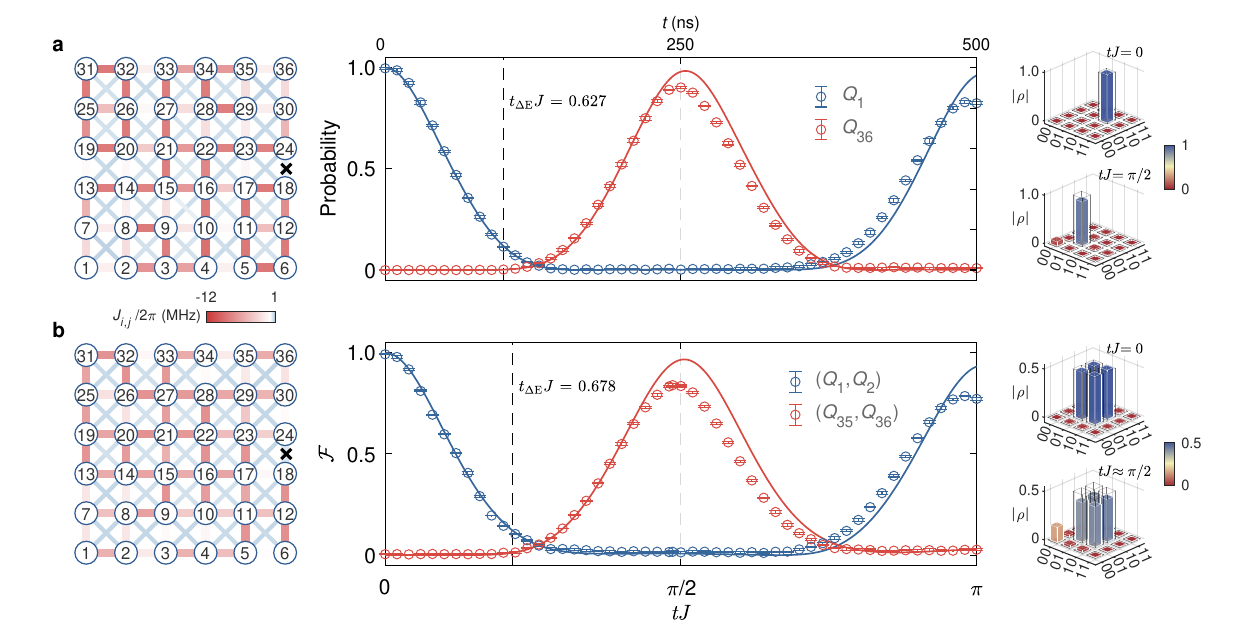}
    \caption{
    {\bf Single-excitation quantum state transfer in a 2D $6\times6$ qubit network with optimized couplings.} 
    {\bf a}, left, shows the measured couplings of the $6\times 6$ qubit network; center, the corresponding time evolution of $Q_1$ and $Q_{36}$ excited-state populations; right, the quantum state tomography in the subspace of the initial and target qubits, $Q_1$ and $Q_{36}$.
    {\bf b}, Fidelity dynamics for the QST using a Bell state initially encoded in qubits $Q_1$ and $Q_{2}$; here, the quantum state tomography at $tJ=0$ is shown in the ($Q_1, Q_2$) subspace whereas at time $tJ\approx\pi/2$~($J/2\pi=1$ MHz) is shown in $(Q_{35}, Q_{36})$. The fidelity here is a generalization of the probability to the Bell case (see text), where we have two basis states in our initial and final wavefunctions to characterize the QST transfer. Lines~(circles) are the numerical~(experimental) evolution with the measured couplings. Solid bars (gray frames) represent experimental (ideal) values of density matrix elements. Error bars come from the standard deviation of five experimental repetitions. $t_{\Delta E}$ is a minimum time for a perfect QST set by `quantum speed limit' arguments~(see Methods). See Supplementary Section \tcr{2} for the specific values of experimentally measured couplings.}
    \label{fig:3}
\end{figure*}

\begin{figure*}
\includegraphics[width=2.0\columnwidth]{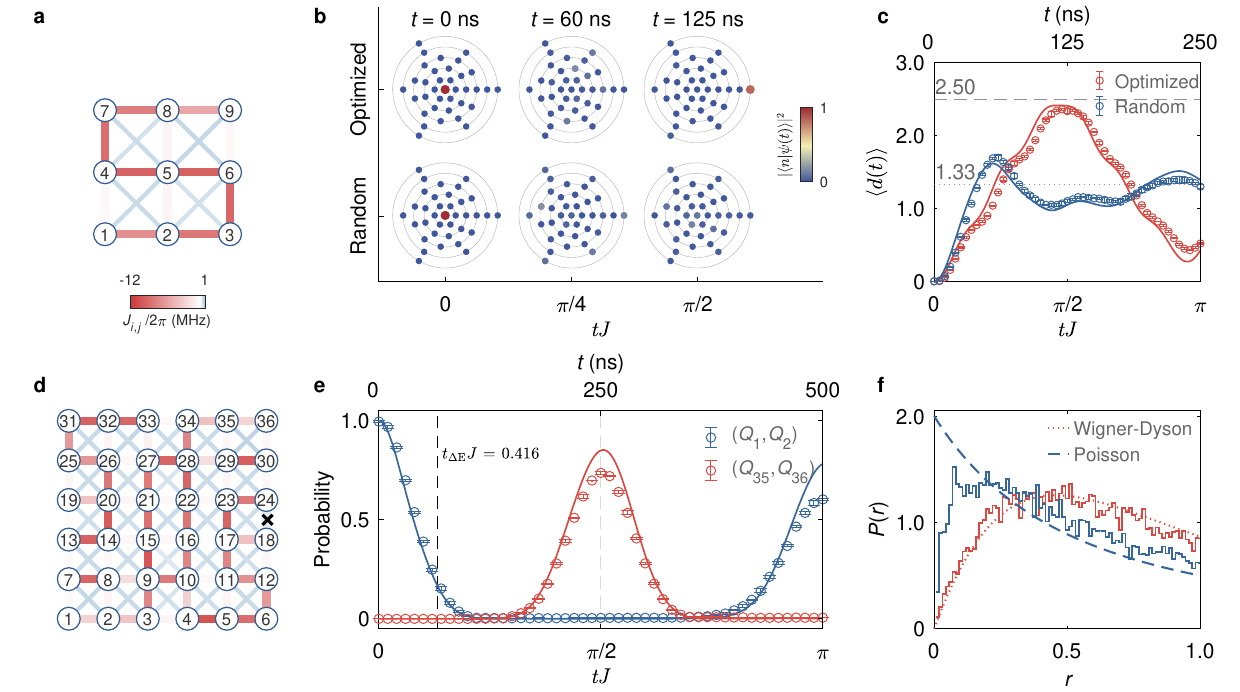}
    \caption{
    {\bf  Two-excitation QST in 2D qubit systems with optimized couplings.} 
    {\bf a,} Measured couplings of the $3\times3$ qubit network for the optimized two-excitation QST (see Supplementary Section \tcr{2} for the specific values). 
    {\bf b,} The corresponding experimental time evolution of two-excitation state QST in Fock space [${\cal D}_{\hat H} = \binom{9}{2} = 36$], where each marker denotes a Fock state - here, we contrast a solution for QST-optimized couplings from one with randomly chosen $J_{ij}$ at different representative times. The concentric circles denote the Fock states with the same distance from the initial state. 
    {\bf c,} The dynamics of the average distance $\langle d(t)\rangle$ traveled in Fock space for both cases; the dashed (dotted) line gives the maximum (mean) distance. 
    {\bf d,} Measured couplings of the $6\times6$ qubit network for the two-excitation QST [${\cal D}_{\hat H}= \binom{36}{2} = 630$] after optimization (see Supplementary Section \tcr{2} for the specific values). 
    {\bf e,} The ($Q_1$,$Q_2$) and ($Q_{35}$,$Q_{36}$) populations over time using the measured couplings in \textbf{d}.  Error bars in \textbf{c} and \textbf{e} come from the standard deviation of five experimental repetitions.
    {\bf f,} Numerically computed distribution of the ratio of adjacent gaps $P(r)$ in the case of QST-optimized and random couplings. Here, we take an average of an ensemble of $k=40$ coupling matrices to improve statistics; dashed and dotted lines are surmises for the Wigner-Dyson and Poisson distributions~\cite{Atas2013}, respectively (see Supplementary Section \tcr{7}). $t_{\Delta E}$ is a minimum time for a perfect QST set by `quantum speed limit' arguments~(see Methods).
    }
    \label{fig:4}
\end{figure*}

Theoretical demonstration of this approach has been put forward in the case of an $N$-site one-dimensional~(1D) $XY$-model quantum spin chain~\cite{Christandl2004, Albanese2004}($\hbar=1$):
\begin{equation}
    \hat H = \sum_{\langle i,j\rangle}^N J_{ij}\, [ \hat \sigma_i^+ \hat \sigma_j^- + \hat \sigma_i^-\hat \sigma_j^+] \ ,
    \label{eq:XX_model}
\end{equation}
where  $\hat \sigma_i^+$ ($\hat \sigma_i^-$) is the raising (lowering) operator for qubit $Q_i$, and the nearest-neighbor (NN) coupling between a pair of qubits is given by $J_{ij}$. The key observation is that provided the couplings are chosen to satisfy $J_{n,n+1} = J \sqrt{n (N-n)}$ for $n=1,\ldots, N-1$, the Hamiltonian~\eqref{eq:XX_model} in the single-excitation subspace is equivalent to that of a large $(N-1)/2$-spin $\vec S$ under the application of a homogeneous magnetic field, i.e., $\hat H/J = \hat S_+ + \hat S_- =  2\hat S_x$, where $\hat S_+~(\hat S_-)$ is the raising (lowering) operator of the large spin. This mapping makes it clear how a perfect QST is realized: An initial state consisting of a single excited qubit $Q_A$, while the remaining qubits are in their lowest states,  $|\psi(0)\rangle = | 1_A 0 0 \cdots 00_B\rangle$, corresponds to the maximal projection of the spin along the $z$-quantization axis in the mapped Hamiltonian. Since the effect of the $x$-oriented uniform field is to precess the large spin, it is evident that at time $tJ = \pi/2$, the state will have full projection along the $-z$-direction. Translating back to the original Hamiltonian, this state corresponds to the reflection-symmetric state $|\psi(tJ = \pi/2)\rangle = | 0_A 0 0 \cdots 01_B\rangle$. Such a perfect QST scheme between opposite qubits $Q_A$ and $Q_B$ in a 1D chain has been previously realized in superconducting quantum circuits featuring four qubits~\cite{Li2018} and photonic qubits in coupled waveguides~\cite{Chapman2016}.

Generalization of these results to higher dimensions is readily obtained in theory~\cite{Kay2010}. For example, in a bipartite lattice in $D$ dimensions, the constraints in the inter-qubit nearest-neighbor couplings satisfy a similar expression: $J_{n,n+1}^{(d)} = J\sqrt{n (N_d-n)}$, for $n=1,\ldots, N_d-1$ and $d = 1,\ldots, D$. The corresponding mapped large-spin Hamiltonian, $\hat H_D/J = 2\hat S_{1,x} + 2\hat S_{2,x}+\ldots +2\hat S_{D,x}$, describes a collection of $D$ large $(N_d-1)/2$ spins, each independently precessing around its $x$-axis at the same rate, thereby guaranteeing perfect QST at time $tJ=\pi/2$. This corollary, derived from the results of Refs.~\cite{Christandl2004, Albanese2004}, has important implications for quantum devices that have increasingly ventured into large-scale two-dimensional superconducting qubit networks~\cite{Arute2019, Xu2023cpl, Wu2021}.

In practice, however, even considering perfectly isolated systems, where decoherence effects are neglected, parasitic cross-couplings~\cite{Zhang2022NP} and device defects~\cite{Arute2019} can naturally occur and, as a result, hamper the perfect QST. The former introduces a connection between qubits across a plaquette via an unwanted coupling $J_{m,m^\prime}^{\times}$ -- see Fig.~\ref{fig:1}{\bf b}. After performing the mapping to the large-spin Hamiltonian,
\begin{eqnarray}
    \hat H_{\rm tot} &=& 2J(\hat S_{1,x} + \hat S_{2,x}) + \nonumber \\ &&\ \ J^{\times}(\hat S_{1,-} \hat S_{2,+}+\hat S_{1,+} \hat S_{2,-} + \hat S_{1,+} \hat S_{2,+}+\hat S_{1,-} \hat S_{2,-}) \nonumber \\
    &=& 2J(\hat S_{1,x} + \hat S_{2,x}) + 4J^{\times} \hat S_{1,x} \hat S_{2,x},
    \label{eq:cross_model}
\end{eqnarray}
such extra terms result in a $J^\times$-mediated spin-spin interaction that spoils the standard predictions~\cite{Christandl2004, Albanese2004} of perfect QST, preventing synchronized precession of the two large spins [Fig.~\ref{fig:1}\textbf{c} and \textbf{d}] (here we assume for simplicity a single energy scale $J^\times$ that governs this term, see Methods). The latter, manifested as a defective coupler in our device [gray bond with a cross marker in Fig.~\ref{fig:1}\textbf{b}], similarly breaks the requirements of NN couplings for a perfect QST.

To overcome these limitations and design a protocol for QST that can be carried over for rather generic situations, we leverage the tuning flexibility of the individual couplings of a 6$\times$6 two-dimensional (2D) superconducting qubit array (see Supplementary Section \tcr{1} for details) and perform a Monte Carlo annealing optimization procedure on the space of parameters of the emulated Hamiltonian such that at a given time $t_{\rm QST}$ one obtains the maximum fidelity for the QST -- details are provided in the Methods Section. That is, we aim to maximize the fidelity of transferring a state $|\psi_A\rangle$  encoded in subsystem $A$ to the target subsystem $B$,  $F(t_{\rm QST}) = |\langle 0_A 0 0 \cdots 0 \psi_B | e^{-{\rm i}\hat H t_{\rm QST}}| \psi_A 0 0 \cdots 00_B\rangle|$,  by finding an optimal solution of the NN coupling parameters $\{J_{n,n+1}^{x\to 1,y\to 2}\}$ that obey the experimental limitations in their tunability and extra constraints arising from imperfections. As a result, one recovers a precession in the mapped large-spin Hamiltonian that is synchronized among the two spins [Fig.~\ref{fig:1}{\bf d}, bottom]. In addition, we adapt our protocol to transfer two excitations across the 2D network, a problem for which analytic treatments are absent. Remarkably, the underlying principle governing a perfect QST of few-particle states is immediately connected with quantum ergodicity and its breaking~[Fig.~\ref{fig:1}{\bf e}]. We experimentally demonstrate these physical insights and realize efficient few-particles transfers in an imperfect $6\times6$ network with fidelities of 0.90 for single-excitation, 0.84 for Bell state, and 0.74 for two-excitation, even if cross-couplings and defect exist. As will become clear in what follows, our work thoroughly dissects few-body state transfer in quantum networks, establishing a unified and fundamental understanding of perfect QST from angular momentum theory and quantum ergodicity. 

\section*{Results}
\subsection*{Single-excitation transfers}
We start by benchmarking the standard one-dimensional protocol of Ref.~\cite{Christandl2004} via employing a single (upper) row of qubits of the current device in Fig.~\ref{fig:1}\textbf{b}, featuring a 1D chain of $N=6$ qubits without discernible cross-couplings [Fig.~\ref{fig:2}{\bf a}]. Figure~\ref{fig:2}\textbf{b} shows a nearly perfect QST with a fidelity above $0.99$ at $t_{\rm QST}\approx125$~ns ($t_{\rm QST}J\approx\pi$/2) by tuning the qubit couplings $J_{n,n+1} = -J \sqrt{n(6-n)}$ (since $\{J_{n,n+1}\}$ are negative in our experiment, we add a minus sign here to keep the label $J$ as a positive value), with $J/2\pi = 2$~MHz (hereafter we define $J$ as a typical energy scale in our experiments with  $J/2\pi = 2$~MHz for 1D and 3$\times$3 cases, and 1 MHz for $6\times6$ cases), which maps onto a single large spin precessing around the $x$-axis [see Fig.~\ref{fig:2}\textbf{c}]. If a single transfer is desired, the remaining qubits can be switched off at $t_{\rm QST}$; otherwise, back-and-forth free propagation of the state occurs between qubits $Q_1$ and $Q_6$, within time scales such that decoherence does not substantially affect device performance. 

To demonstrate the effectiveness of our procedure, we begin with a case where a defective coupling is deliberately included, and the resulting QST fidelity is low. For that, we explore quantum information transfer in a subset of qubits in Fig.~\ref{fig:1}\textbf{b}: a $3\times 3$ 2D network with its lower left corner, $Q_{4, 2}$ [see Supplementary Fig.~\tcr{S1}], relabeled as $Q_1$ in Fig.~\ref{fig:2}\textbf{d} and \textbf{g}. It encompasses a device defect -- a malfunctioning coupler [the bond with a cross marker in Fig.~\ref{fig:2}\textbf{d} and \textbf{g}] that constrains one of the qubit couplings to $\sim2\pi \times 0.3$~MHz. NN couplings are parametrically calibrated with $J_{n,n+1}^{(1, 2)} = -J\sqrt{n(3-n)}$ on other qubit pairs [Fig.~\ref{fig:2}{\bf d}]. Under these conditions, Figure~\ref{fig:2}{\bf e} displays the results of the excited-state population dynamics for $Q_1$ and $Q_9$, quantifying the transfer between a single-qubit excitation initialized at $Q_1$ and aimed to transfer it to the opposite qubit $Q_9$. Unfortunately, the transfer success is largely compromised to a low fidelity of about $0.27$ precisely because the existing  $J^{\times}_{m,m^{\prime}}$-couplings and the defect prevent the standard coupling parametrization~\cite{Christandl2004} from achieving perfect QST. In the language of the mapped Hamiltonian, the rotations of the two spins are now correlated, leading to the failure of approaching the pole of the $-z$ direction -- see experimental and simulated results in Fig.~\ref{fig:2}{\bf f} (Supplementary Section \tcr{5} details the measurements of trajectories). 

Having shown that parasitic and defective couplings in real devices destroy the transmission of a quantum state with the standard protocol~\cite{Christandl2004}, we now tackle these limitations by careful tuning (see Supplementary Section \tcr{2}) of the coupler-mediated interactions $J_{n,n+1}^{(1,2)}$ [Fig.~\ref{fig:2}{\bf g}] according to the couplings optimized with the aforementioned annealing optimization procedure (see Supplementary Section \tcr{8}). Figure \ref{fig:2}{\bf h} reports these results for the same $3\times 3$ network: Regardless of cross-couplings and one fixed coupling, the transfer fidelity is greatly improved, reaching a value of $0.936\pm0.012$. In such a scenario, the trajectories of two coupled large spins $\{\langle\hat S_{(1,2),x}(t)\rangle,\langle\hat S_{(1,2),y}(t)\rangle,\langle\hat S_{(1,2),z}(t)\rangle\}$, are optimized in a way to reach the $-z$ poles synchronously despite their different paths during the dynamics [see Fig.~\ref{fig:2}\textbf{i}]. These results pave the way for pursuing QST in much larger quantum circuits, where imperfections are more likely to occur~\cite{Arute2019, Xu2023cpl}.

By employing all 36 qubits and utilizing the optimization procedure under those constraints (see Methods), we report in Fig.~\ref{fig:3}{\bf a} the transfer of a single excitation across a $6\times6$ qubit network. Here we experimentally achieve a maximum transfer fidelity of $0.902\pm 0.006$~[see Fig.~\ref{fig:3}\textbf{a}]. Experimentally reconstructed density matrices, labeled by  $\rho$, of the initial state in $Q_1$ and the resulting final state in $Q_{36}$ after QST are shown in the right panels of Fig.~\ref{fig:3}\textbf{a}. Under ideal conditions, we can numerically obtain solutions for optimized NN couplings with QST-fidelities above 0.99, even if influenced by cross-couplings and defects (see Supplementary Section \tcr{8}), but experimental imperfections in calibrating couplings and qubit frequencies can impact those results. More prominent, however, are the residual thermal excitations in the qubit network, which mainly cause the observed experimental infidelity. Numerical simulations suggest that $0.5\%$ thermal excitations in each qubit could result in transfer errors of $\sim3\%$ for $3\times3$ network and $\sim10\%$ for a $6\times6$ network. This indicates that their suppression constitutes an important route to further improve the future transfer fidelity on a 2D network. For a detailed discussion on the effects of thermal excitations and the noise analysis of couplings and qubit frequencies, see Supplementary Sections \tcr{3} and \tcr{4}, respectively.

Building on these results, our protocol similarly accomplishes the QST of maximally entangled two-qubit states. By preparing a Bell state $|\Psi^{-}\rangle = \big(\, |01\rangle - |10\rangle \, \big) / \sqrt{2}$ in qubit pair $(Q_{1}, Q_{2})$, we target the transfer to the opposite-symmetric qubit pair $(Q_{35}, Q_{36})$ in the network. For that, the initial state $|\Psi^{-}\rangle$ is obtained by applying a quantum circuit, which consists of a two-qubit control-Z gate and several single-qubit gates, on $Q_1$ and $Q_2$ (see Supplementary Section \tcr{6}). After a transfer time $tJ\approx\pi/2$, we perform two-qubit quantum state tomography on $Q_{35}$ and $Q_{36}$ to witness the transfer efficiency and find a fidelity $\mathcal{F}={\rm tr}(\rho_{\rm exp}\rho_{\rm ideal})\approx$ 0.840$\pm$0.006 for the experimentally reconstructed density matrix $\rho_{\rm exp}$, which demonstrates the effectiveness of our protocol for transferring quantum entanglement. Here, the QST fidelity of the entangled state displays a large sensitivity to noise in both the qubit's frequency and the value of the optimized couplings (see Supplementary Section \tcr{4} for an extended discussion), which substantially restricts the transfer success. 

\subsection*{Two-excitation transfers}
The observed relatively large transfer fidelity for states (entangled or not) composed of a single excitation endows the ability to push toward an even more challenging scheme. A standard mapping (see Methods) of the spin-operators in~\eqref{eq:XX_model} relates the emulated model to one of hardcore bosons, whose cardinality reflects the number of photon excitations and hopping energies $J_{ij}$. In the case of a two-dimensional lattice, such a model describes a typical quantum chaotic Hamiltonian. Its non-integrability renders quick thermalization and the absence of memory of the initial conditions throughout sufficiently long dynamics~\cite{Rigol2008}. Therefore, the prospects of achieving a successful QST are slim: One expects diffusive behavior in Fock space~\cite{yao2023np}, making it unlikely to find a single state with the majority of the weight in $|\psi(t)\rangle$ at a later time. Yet, suppose the number of excitations is small compared to the total system size. In that case, we argue that this weakly chaotic Hamiltonian~\cite{Schiulaz2018, Fogarty2021} can still be engineered such that a two-excitation QST is efficient (see Supplementary Section \tcr{8} for the optimized couplings).

Figure~\ref{fig:4}\textbf{a} and \ref{fig:4}\textbf{d} show the QST-optimized couplings for two-excitation states in $3\times 3$ and $6\times 6$ qubit networks -- in both cases, the pair $(Q_{1}, Q_{2})$ is initially excited. In the former, we experimentally exemplify in Fig.~\ref{fig:4}\textbf{b} the propagation of $|\psi(t)\rangle$ in Fock space, contrasting both the optimized and random couplings at different times. Only when the couplings are optimized does one recover a regime where most of the weight collapses on a single Fock state, quantitatively describing the schematic cartoon in Fig.~\ref{fig:1}{\bf e}. The states are organized according to a metric defined by the $L_1$-norm of the excitations in the lattice (see Methods) such that the distance $d$ to the initial state $|n=0\rangle$ obeys $d(|0\rangle,|0\rangle) = 0$, whereas the distance to the target state, $d(|0\rangle,|n_{\rm target}\rangle)$, takes the maximum value for the given network size. 
Thus, an average distance can be dynamically defined as
\begin{equation}
    \langle d(t)\rangle = \sum_{n=0}^{{\cal D}_{\hat H}-1} d(|0\rangle, |n\rangle)\  |\langle n|\psi(t)\rangle|^2\ ,
\end{equation}
measuring the wave-packet's `center of mass' evolution in the Fock space of dimension ${\cal D}_{\hat H}$. Figure~\ref{fig:4}{\bf c} displays its dynamics for the case of the QST-optimized solution of the couplings, where one observes a ballistic (almost) periodic evolution between the initial state $|0\rangle$ and the target state $|n_{\rm target}\rangle$. Conversely, if one implements random couplings $\{J_{ij}\}$ between the qubits, a slow evolution towards the mean distance $\overline{d} = \frac{1}{\cal D_{\hat{H}}}\sum_{n=0}^{{\cal D}_{\hat H}-1} d(|0\rangle, |n\rangle)$ ($\overline{d}\approx1.33$ for 3$\times$3 case) is achieved. In this case, the wave-packet dynamics after an initial transient is close to exhibiting a diffusive behavior  (Supplementary Section \tcr{7}).

Turning to the large $6\times 6$ qubit network, we report in Fig.~\ref{fig:4}{\bf e} the population dynamics of the initial and target two-excitation states: Here, a maximum fidelity of about $0.737\pm 0.007$ is experimentally observed for the transfer of excitations from $(Q_{1}, Q_{2})$ to $(Q_{35}, Q_{36})$. Notwithstanding the large Hilbert space ${\cal D}_{\hat H} = \binom{36}{2}=630$ one has to deal with in this case to find the optimized couplings that maximize the QST fidelity, the non-integrability of the Hamiltonian naturally bounds performance. Minor deviations on the optimized couplings, which can inherently occur owing to experimental imperfections, significantly impact the transfer fidelity~(see Supplementary Section \tcr{4}).

The approach we have introduced here explicitly uses an annealing Monte Carlo procedure to optimize QST.  As such, it is a `black box', providing no clear physical indication of \textit{why} the optimized coupling solutions give a better QST.
Indeed, there is no discernible pattern in the optimized $J_{ij}$, unlike the case of the 1D protocol. To acquire that insight, we perform an ergodicity analysis, classifying the eigenspectrum $\{\varepsilon_\alpha\}$ of the corresponding Hamiltonians using the ratio of adjacent gaps $r_\alpha \equiv \min(s_\alpha,s_{\alpha+1})/\max(s_\alpha,s_{\alpha+1})$~\cite{Oganesyan2007}, where $s_\alpha = \varepsilon_{\alpha+1} - \varepsilon_\alpha$. If the couplings are randomly chosen, Fig.~\ref{fig:4}{\bf f} shows that the distribution $P(r)$ typically follows one of the random matrices of the same symmetry class of $\hat H$, signifying strong ergodicity in the spectrum where level repulsion takes place $[P(r=0) \to 0]$ (see Supplementary Section \tcr{7} for an extended discussion). In contrast, if the couplings are optimized, ergodicity is largely absent, and a distribution $P(r)$ close to a Poisson one is obtained instead. This analysis thus provides a clear understanding of the underlying interpretation of the optimization procedure: the couplings $J_{ij}$ evolve to partially cure the quantum chaotic nature of the system, allowing higher fidelity QST to take place.

Our Monte Carlo annealing process does not produce a unique solution. Different $J_{ij}$ can be found which give QST of high fidelity. The ergodicity analysis provides a link between these solutions: They share a  distribution of their level spacings, which is roughly Poissonian, with higher fidelities linked to more faithful Poisson statistics.

\section*{Discussion}
We demonstrate highly efficient QST of few-excitation states, even with natural imperfections, in manufactured superconducting quantum circuits. The key ingredient relies on manipulating the inter-qubit couplings, whose values are set by a classical optimization procedure under the conditions of maximization of the fidelity of the transfer at a given time $t_{\rm QST}$. This time scale is bound to obey two constraints: If $t_{\rm QST}$ is large, typical couplings are small in magnitude, but the longer the time, the more drastic are the effects of decoherence, which would ultimately inhibit an efficient quasi-adiabatic QST. On the other hand, a fast QST is bounded by fundamental limits of the evolution of \textit{any} quantum mechanical system. Dubbed quantum speed limits, they control the minimal time scale necessary for a time-evolving wave function to become fully distinguishable (i.e., orthogonal) to the initial state. Such a perfect orthogonalization process precisely describes a flawless QST. Known bounds~\cite{Mandelstam1944, Margolous1998} limit the minimal orthogonalization time based on either the mean energy $E = \langle \hat H\rangle$ or the energy uncertainty $\Delta E = \sqrt{\langle\hat{H}^2\rangle-\langle\hat{H}\rangle^2}$ of the system
\begin{equation}
    t_{\tiny \Delta E}=\frac{\pi\hbar}{2\Delta E}\ \ \ \ \ \ t_{\tiny E}=\frac{\pi\hbar}{2(E-E_g)}\ ,
    \label{eq:t_qsl}
\end{equation}
with $E_g$ the ground-state energy of the Hamiltonian $\hat H$ that governs the unitary dynamics. As a result, $t_{\rm QST} \geq \max\{t_{\tiny \Delta E}, t_{\tiny E}\}$. In Figs.~\ref{fig:2}, \ref{fig:3}, and \ref{fig:4}, we include the minimal orthogonalization time given by the quantum speed limit for each of the Hamiltonians that describes the corresponding evolution (see Supplementary Section \tcr{9} for an extended analysis). In all cases we investigate, the energy uncertainty limit bounds the QST, i.e., $t_{\rm QST}\geq t_{\tiny \Delta E}$.

Our results build a firm understanding of the efficient transfer of few-particle states in 2D networks, even in the presence of unwanted parasitic couplings and inherited defects. These demonstrations on an actual large-scale physical device underline the significance of our protocol, not only in implementing real-world quantum communication channels or distributing entanglement across the solid-state device but also in providing a constructive technique for utilizing engineered quantum networks as building blocks for further quantum applications.

\noindent{\bf Data availability}\\
The data presented in the figures and that support the other findings of this study will be publicly available upon its publication.

\noindent{\bf Code availability}\\
The codes used for numerical simulation are available from the corresponding authors upon reasonable request.

\noindent {\bf Acknowledgments}\\
We thank Siwei Tan and Liangtian Zhao for their technical support.
RM acknowledges insightful discussions with David Weiss, Marcos Rigol, and Pavan Hosur. The device was fabricated at the Micro-Nano Fabrication Center of Zhejiang University. This research was supported by the National Natural Science Foundation of China (Grants No. 92065204,NSAF-U2230402, 12222401, 11974039, U20A2076, 12274368), the Zhejiang Province Key Research and Development Program (Grant No. 2020C01019). RTS was supported by the grant DE‐SC0014671 funded by the U.S.~Department of Energy, Office of Science. QG is also supported by Zhejiang Provincial Natural Science Foundation of China (Grant Nos. R24A040010, LQ23A040006) and Zhejiang Pioneer (Jianbing) Project (Grant No. 2023C01036). 

\noindent{\bf Author contributions}\\
R.M. and R.T.S proposed the idea;
A.Y., J.P., Z.Z., and R.M. performed the numerical simulations;
L.X. and Z.Z. conducted the experiment under the supervision of Q.G. and H.W.;
J.C. and H.L. designed and fabricated the device under the supervision of H.W.;
R.M., Q.G, and R.T.S. co-wrote the manuscript;
and all authors contributed to the experimental setup, discussions of the results, and manuscript development.  

\noindent{\bf Competing interests}\\
The authors declare no competing interests.

\noindent{\bf Additional information}\\

\section*{Methods}
\subsection*{Monte Carlo annealing process}
The task of accomplishing an efficient quantum state transfer relies on an optimization scheme to find appropriate coupling matrices $\{J_{m,n}^{x,y}\}$ for NN qubits $Q_m$ and $Q_n$, in the presence of intraplaquette couplings $J^{\times}_{m,m^\prime}$ and the defect. For that, one employs a Monte Carlo process in this space of parameters with cost function $p({\{J_{m,m+1}^{x,y}, J^\times_{m,m^\prime}\}}) \propto e^{-\tilde F/T}$, where the `temperature' $T$ is varied from $T_{\rm high}\to T_{\rm low}$ and $\tilde F$ marks the quantity one aims to minimize: the infidelity $\tilde F(t_{\rm QST}) = 1 - |\langle \psi(t_{\rm QST})|\psi_{\rm target}\rangle|^2 $ of the perfect quantum state transfer at times $t_{\rm QST}$. At each step of the sampling, one obtains such state via unitary time-evolution $|\psi(t_{\rm QST})\rangle = e^{-{\rm i}\hat H t_{\rm QST}}|\psi(0)\rangle$, where $\hat H$ is constructed with the current couplings parameters $\{J_{m,m+1}^{x,y}, J^\times_{m,m^\prime}\}$. Throughout the sampling, we use a combination of local and global parameter changes, combined with $k$ independent realizations of the Monte Carlo process ($k = 40$ for the two-excitation transfer and $k=5$ for the remaining cases). Furthermore, we also compare different annealing scheduling protocols $f(T)$, where $f(T_{\rm high(low)}) = T_{\rm high(low)}$, and proposed changes in the couplings are dynamically adjusted according to their acceptance ratio.

Among the many choices for inter-qubit coupling matrices that can maximize the fidelity of quantum state transfer, we focus on the ones that preserve the network inversion symmetry when dealing with single- or two-excitation transfers. Two reasons stand behind this: (i) such symmetry is also present in the original protocols for perfect quantum state transfer~\cite{Christandl2004}, and (ii) it reduces the space of parameters one needs to probe to find solutions that minimize the infidelity. Under such conditions, we must optimize 30 individual couplings (15 for each direction) on the $6\times 6$ superconducting quantum circuit instead of 60 couplings if no symmetries were enforced.

Over an extensive experimental calibration process (see Supplementary Section \tcr{2}), we properly approximate all the next-NN coupling parameters (parasitic couplings) to a value of $J^\times_{m,m^\prime}/2\pi=0.45$ MHz in the optimization. As a result, the optimization process proceeds with this extra constraint. Lastly, as pointed out in the text, the coupler connecting qubits $Q_{18}$ and $Q_{24}$ is defective, setting the corresponding coupling $J_{18,24}/2\pi$ to a fixed value $+0.3$ MHz. This is similarly taken into account in the annealing process.

Finally, we remark that in a sufficiently large lattice, a possible solution for good QST for multiple excitations is to have disjoint paths along which individual excitations propagate independently.  However, we have verified that more complex solutions with equally high fidelity also exist by forcing all the couplings $J_{ij}$ to be bounded away from the origin. Typically, we have performed the annealing process enforcing that $\{J_{m,n}^{x,y}\}/2\pi \in [J_{\rm min}, J_{\rm max}]$ MHz. In general, a larger range yields a higher optimized QST fidelity but becomes more challenging for experimental calibrations. In practice, we set different $J_{\rm min} \in [-12, -6]$ and $J_{\rm max} \in[-0.5, -0.3]$ for different cases (see Supplementary Section \tcr{8} for the details of optimized couplings and QST dynamics); ergodicity analysis in Fig.~\ref{fig:4}{\bf f} is performed with $\{J_{m,n}^{x,y}\}/2\pi \in [-10, -0.1]$ MHz.

\subsection*{Cross-couplings}
When using the functional form that maximizes QST in a regular lattice, mapping it to a large-spin Hamiltonian, $J_{n,n+1}^{(d)} = J\sqrt{n (N_d-n)}$, possible cross-couplings among the qubits can be incorporated in a similar picture such that $J^\times_{m,m^\prime} = J^\times\sqrt{m (N_1-m)}\sqrt{m^\prime (N_2-m^\prime)}$, with $m=1,\ldots,N_1-1$ and $m^\prime=1,\ldots,N_2-1$. This is assumed when writing the compact Hamiltonian in Eq.~\ref{eq:cross_model}. Experimentally, however, active control of these couplings is inaccessible, and as mentioned above, calibration shows that $J^\times_{m,m^\prime}$ is approximately constant over the qubit network. As a result, while the final emulated Hamiltonian evades the simple large-spin form in this situation, one can still generically represent and compute the corresponding spin `trajectories,' as done in Fig.~\ref{fig:1}{\bf d}, even in the presence of non-parameterized cross-couplings.

\subsection*{Hardcore boson picture}
We use the notation of qubit excitations and particles in the text interchangeably. This relies on the standard mapping between hardcore bosons and the spin-1/2 operators: $\hat a_i^\dagger \leftrightarrow \hat \sigma_{i}^+$ and $\hat a^{\phantom{\dagger}}_i \leftrightarrow \hat \sigma_{i}^-$~\cite{Matsubara1956}. As a result, the Hamiltonian \eqref{eq:XX_model} is written as
\begin{align}
    \hat H = \sum_{\langle i,j\rangle}^N J_{ij} [ \hat a_i^\dagger \hat a^{\phantom{\dagger}}_j + \hat a_j^\dagger\hat a^{\phantom{\dagger}}_i] \ ,
\end{align}
where the `couplings' are read as hoppings energies between orbitals $i$ and $j$.

\subsection*{Distances in Fock space}
Given the typical values of the coupling's matrix, we define a metric for distances between Fock states inspired by the associated time for a Fock state to be reached. For example, the target state with excitations in the qubit-pair $(Q_{35}, Q_{36})$ in the $6\times 6$ qubit network should be one of the most distant from the initial state with excitations in $(Q_{1}, Q_{2})$. Using the initial state as a reference, a possible distance is defined as $d(|0\rangle,|n^\prime\rangle) = \frac{1}{4}\sum_{l=1}^2\left(|x_l^\prime - x_0| + |y_l^\prime - y_0| + |x_l^\prime - x_1| + |y_l^\prime - y_1|\right) - 1/2$, where $(x_l^\prime,y_l^\prime)$ are the cartesian coordinates of each of the $l$-excitations ($l=2$) of a generic Fock state $|n^\prime\rangle$. For the initial state $|n=0\rangle$, one thus have $(x_0,y_0)$ and $(x_1,y_1)$ being the coordinates of its excitations. Hence $d(|0\rangle,|0\rangle) = 0$ whereas $d(|0\rangle,|n_{\rm target}\rangle) = 8.5 
 ~(2.5)$ for the target state in the $6\times 6~(3\times 3)$ network size. The $1/4$ prefactor in the definition of $d(|0\rangle,|n^\prime\rangle)$ refers to the average of the four different $L_1$-norm distances to each pair of particles in the two Fock states, $|0\rangle$ and $|n^\prime\rangle$, owing to the particle's indistinguishability.

\bibliography{QST.bib}

\end{document}


\nolinenumbers
\title{Supplementary materials for
\\Enhanced quantum state transfer: Circumventing quantum chaotic behavior
}
\maketitle
\onecolumngrid

\tableofcontents
\beginsupplement
\section{Device information}

\begin{figure*}[h]
  \includegraphics[width=0.9\columnwidth]{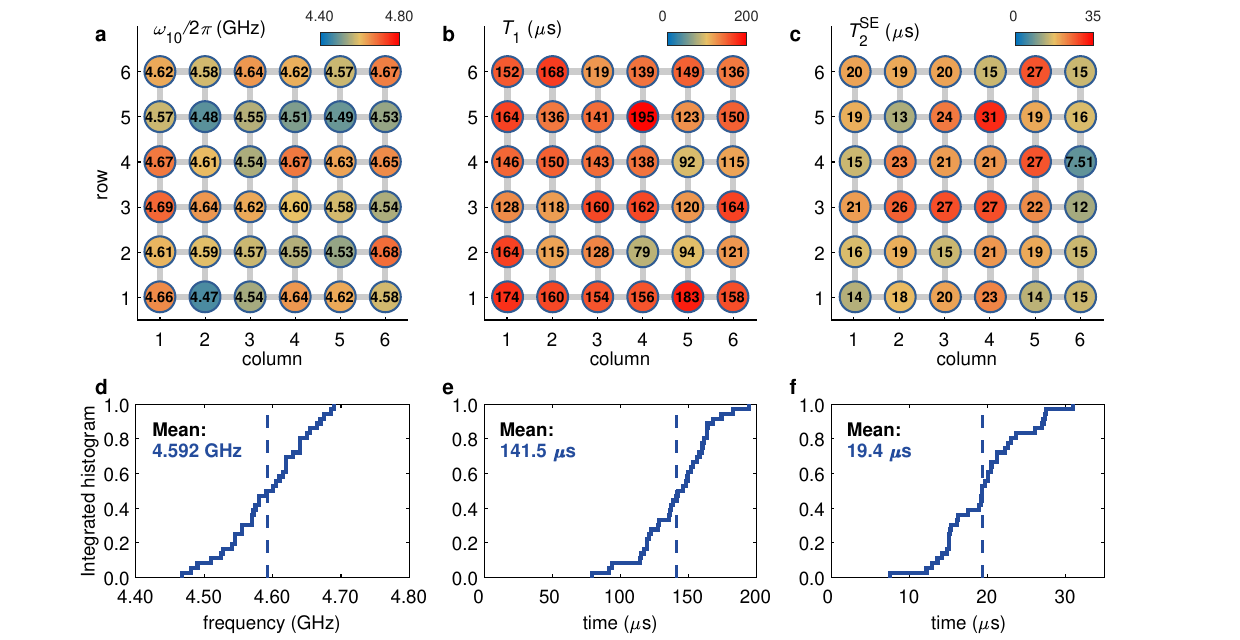}
   \vspace{-0.4cm}
   \caption{{\bf Coherence time for 36 qubits.}
   \textbf{a - c}, Heat map for typical values of idle frequency $\omega_{\rm 10}$ (\textbf{a}), energy relaxation time $T_{\rm 1}$~(\textbf{b}), and spin-echo dephasing time $T^{\rm SE}_2$~ (\textbf{c}). $T_1$ and $T^{\rm SE}_2$ are measured near interaction frequency $\omega_I/2\pi=4.585$ GHz where quantum state transfer happens.
   \textbf{d - f}, Histogram statistics of the corresponding results from  \textbf{a - c}, where the mean value of each panel is highlighted.
   }
   \label{subfig:info_coherence}
\end{figure*}

A two-dimensional (2D) flip-chip superconducting quantum processor is utilized for our quantum state transfer~(QST) experiments, which includes $36$ qubits and $60$ couplers. These qubits are built as a $6\times6$ square lattice, where a tunable coupler connects each nearest-neighbor (NN) qubit pair. Each qubit has an individual control line for single-qubit rotations and frequency modulation via applying microwave (XY) pulses and flux (Z) pulses. Each coupler also has a control line for tuning the coupling strength of the connected qubit pair by applying the flux pulses. Details of this processor's experimental setup for wiring and control electronics can be found in Ref.~\cite{yao2023np}.

\begin{figure*}[h]
  \includegraphics[width=0.9\columnwidth]{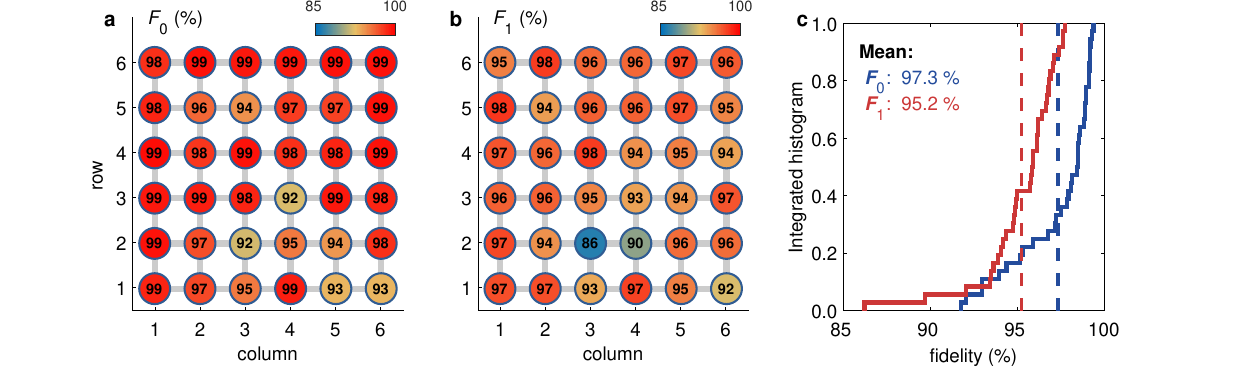}
   \vspace{-0.4cm}
   \caption{{\bf Readout performance for 36 qubits.}
   \textbf{a}, Heat map of readout fidelity $F_{\rm 0}$, which is obtained by preparing all the qubits in their ground states and measuring the probability of $|0\rangle$ state for each qubit.
   \textbf{b}, Heat map of readout fidelity $F_{\rm 1}$. Each value is measured by only exciting the target qubit and detecting the corresponding $|1\rangle$ state probability.
   \textbf{c}, Histogram statistics of $F_{\rm 0}$ and $F_{\rm 1}$, where the mean value of each panel is highlighted.
   }
   \label{subfig:info_read}
\end{figure*}

Up to $36$ qubits and $59$ couplers are dynamically controlled during the experiments since the coupler between $Q_{18}$ and $Q_{24}$ malfunctions and presents limited control. Figures~\ref{subfig:info_coherence} and \ref{subfig:info_read} display typical performance for all 36 qubits, such as qubit idle frequency ${\omega}_{10}$, energy relaxation time $T_1$, spin-echo dephasing time $T^{\rm SE}_2$, and readout fidelities $F_0$ and  $F_1$. Notably, the average energy relaxation time $T_{1}$ near interaction frequency ($\omega_I/2\pi=4.585$ GHz) is above 140 $\mu$s (see Figs.~\ref{subfig:info_coherence}{\bf b} and {\bf e}), and the spin-echo dephasing time $T^{\rm SE}_2$ has the mean value of $\sim$ 19 $\mu$s (see Figs.~\ref{subfig:info_coherence}{\bf c} and {\bf f}). Readout fidelities (see Fig.~\ref{subfig:info_read}) are measured individually for each qubit, assisted with a reset protocol to reduce thermal populations.

\section{Measurements of coupling strengths}

\begin{figure*}[h]
\vspace{-0.5cm}
\includegraphics[width=1.0\columnwidth]{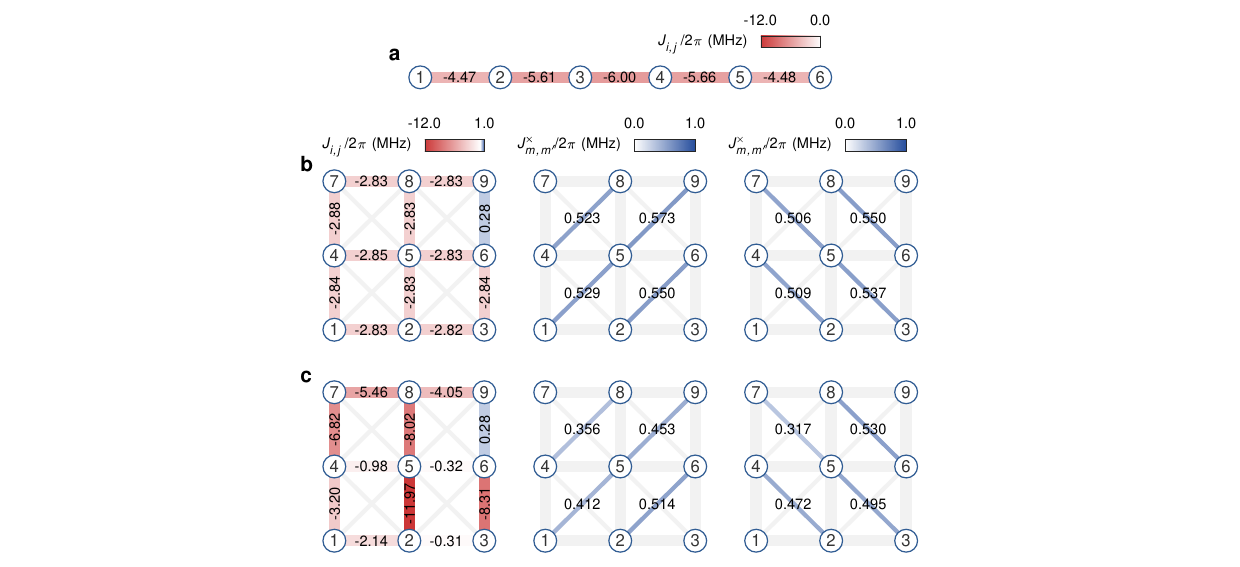}
\vspace{-0.7cm}
\caption{\textbf{Experimentally measured couplings for single-excitation QST in small systems.}
    \textbf{a}, Measured couplings for single-excitation QST on a  6-qubit 1D chain.  This coupling configuration is the standard 1D protocol of Ref.~\cite{Christandl2004} and the corresponding dynamics are shown in Figs.~\tcr{2}\textbf{b} and \textbf{c} of the main text.
    \textbf{b}, Measured couplings for a $3\times3$ network with NN couplings parameterized using the standard protocol of Ref.~\cite{Christandl2004}.  In the ideal case, all the NN couplings are expected to be homogeneous for the $3\times3$ network,  but due to the defect between $Q_6$ and $Q_9$, this coupling is fixed with $J_{\rm defect}/2\pi \approx 0.3$ MHz. The corresponding transfer dynamics are shown in Figs.~\tcr{2}\textbf{e} and \tcr{2}\textbf{f} of the main text.
    \textbf{c}, Measured couplings optimized with largely successful QST (see main text, Figs.~\tcr{2}{\bf h} and {\bf i}).}
    \label{subfig:Couplings_1x6and3x3}
\end{figure*}

In our protocol, the nearest-neighbor couplings and the residual cross-couplings provide channels for quantum state transfer, which makes their accurate measurements fundamental for accomplishing the desired effects. The coupling strength between each qubit pair is obtained via resonant photon swap dynamics. For each NN qubit pair, we measure its coupling strength twice, with all other qubits detuned $\pm\Delta$  away from the target interaction frequency~($\omega_I$), and all other NN couplings kept at the same values as those in the QST process. Then, we can estimate the absolute value of the coupling strength by averaging over the two measurement results. Since NN couplings are dominated by the virtual photon exchange mediated by the coupler, in the regime we consider, if the coupler's frequency is above the interaction frequency, the sign is negative; otherwise, it is positive.

\begin{figure*}[h]
\includegraphics[width=1.0\columnwidth]{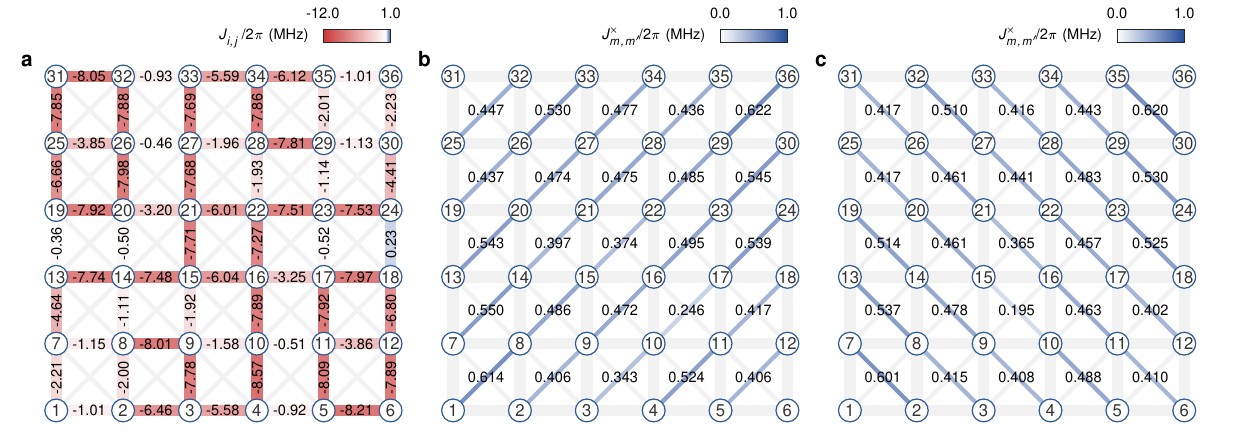}
    \caption{\textbf{Experimentally measured couplings for single-excitation QST in the $6\times6$ network.} 
    \textbf{a},  Measured NN couplings including the defect between $Q_{18}$ and $Q_{24}$, utilizing an annealing optimized solution as target -- see Fig.~\tcr{3}{\bf a} in the main text.  The defective coupling is about $0.2$ MHz.
    \textbf {b} and \textbf {c}, Measured cross couplings.  
    }
    \label{subfig:Couplings 6x6 single excitation}
\end{figure*}

\begin{figure}[htp]
\vspace{-0.7cm}
\includegraphics[width=1.0\columnwidth]{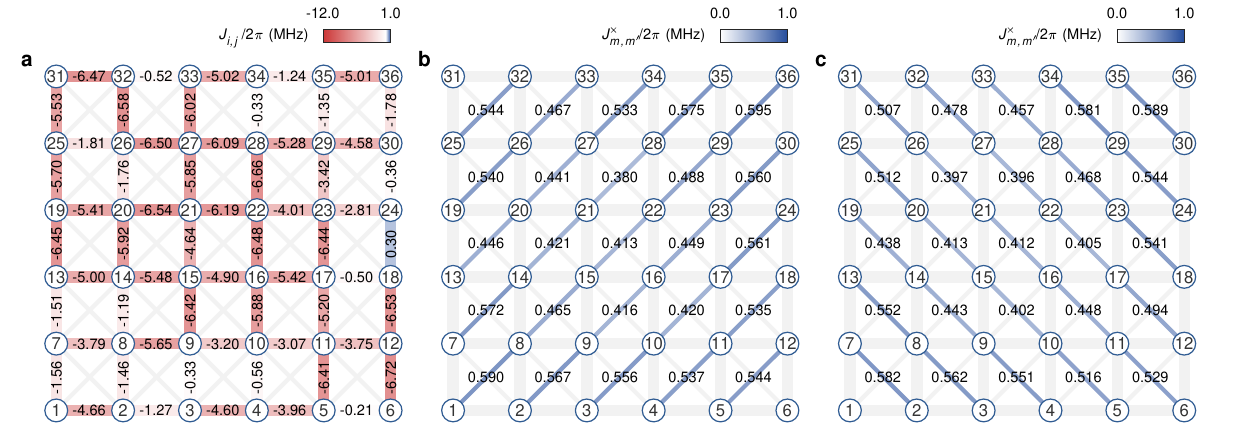}
\vspace{-0.7cm}
    \caption{\textbf{Experimentally measured  couplings for Bell state QST in the $6\times6$ network.} 
    \textbf{a}, Measured NN couplings including the defect between $Q_{18}$ and $Q_{24}$, utilizing an annealing optimized solution as target -- see Fig.~\tcr{3}{\bf b} in the main text. The defective coupling is about $0.3$ MHz.
    \textbf{b} and \textbf{c}, Measured cross-couplings.} 
    \label{subfig:Couplings 6x6 Bell state}
\end{figure}

\begin{figure}[htp]
\includegraphics[width=1.0\columnwidth]{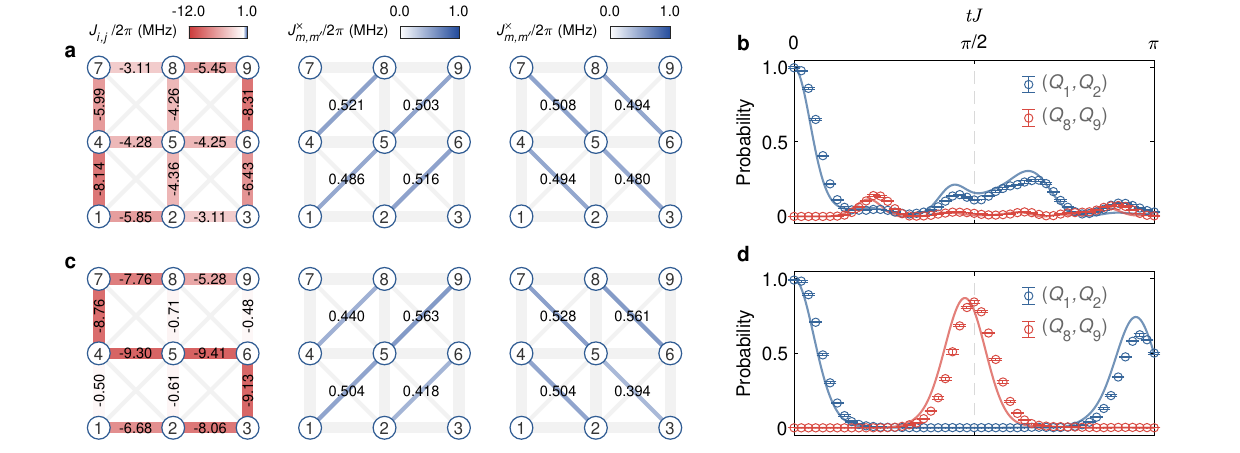}
\vspace{-0.7cm}
    \caption{\textbf{Experimentally measured couplings for two-excitation QST in the $3\times3$ network.} 
    \textbf{a}, Measured NN  and cross-couplings for the two-excitation QST in a $3\times3$ network with randomly selected NN couplings.
    \textbf{b},  Experimentally measured population dynamics (markers) and numerical results (lines) using the random coupling configuration in \textbf{a}.
    \textbf{c}, Measured NN  and cross-couplings for the enhanced two-excitation QST in a $3\times3$ network with the annealing optimized NN couplings.   
    \textbf{d},  Experimentally measured population dynamics (markers) and numerical results (lines) with the optimized NN couplings in \textbf{c}. The corresponding experimental results for the dynamics in Fock space are shown in Figs.~\tcr{4}{\bf b} and {\bf c} of the main text.} 
    \label{subfig:Couplings 3x3 two excitation}
\end{figure}

\begin{figure}[htp]
\includegraphics[width=1.0\columnwidth]{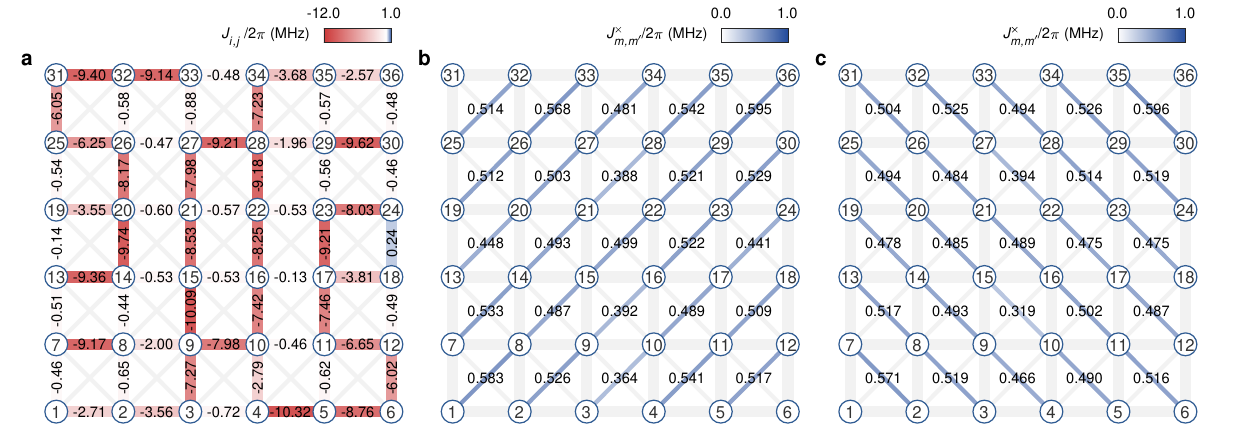}
\vspace{-0.7cm}
    \caption{\textbf{Experimentally measured couplings for two-excitation QST in the $6\times6$ network.} 
    \textbf {a}, Measured NN couplings. The defect between $Q_{18}$ and  $Q_{24}$ is $J_{\rm defect}/2\pi\approx 0.2$ MHz.
    \textbf {b} and \textbf{c}, Measured cross-couplings. The corresponding experimental dynamics for transferring two excitations from $(Q_1,Q_2)$ to $(Q_{35},Q_{36})$ are shown in Fig.~\tcr{4}{\bf e} of the main text. 
    }
    \label{subfig:Couplings 6x6 two excitation}
\end{figure}

The measurements are similar but relatively more complex for cross-coupling $J^{\times}$. We find that the measured coupling, labeled by $J(\Delta)$, depends on both NN coupling configurations and the frequencies of other qubits. Therefore, we measure $J(\Delta)$ for a set of detunings $\{ \Delta_i\}$, such as $\Delta_i/2\pi=-150~{\rm MHz}, -120~{\rm MHz}, -100~{\rm MHz}, 150~{\rm MHz}$, and then estimate $J^{\times}$'s value by fitting the results with an approximate formula $J({\rm \Delta}) = J^{\times} + {\rm sign}(g)\frac{2g^2}{\Delta}$. Here, $g$ is a factor representing the effective averaged NN couplings in the same plaquette, and $J^{\times}$ is the cross-coupling we want to measure. Physically, $J^{\times}$ is mainly caused by a parasitic capacitor directly connecting the two diagonal qubits in the plaquette. In addition, the weak coupling between the qubit and its next NN coupler also leads to a small contribution to $J^{\times}$ via a virtual photon interaction. 
Experimentally measured couplings for different cases of QST highlighted in the main text are shown in Fig.~\ref{subfig:Couplings_1x6and3x3}~(single-excitation QST in $1\times6$ and $3\times3$ networks), Fig.~\ref{subfig:Couplings 6x6 single excitation}~(single-excitation QST in $6\times6$ network), Fig.~\ref{subfig:Couplings 6x6 Bell state}~(Bell state QST in $6\times6$ network), Fig.~\ref{subfig:Couplings 3x3 two excitation}~(two-excitation QST in $3\times3$ network), and Fig.~\ref{subfig:Couplings 6x6 two excitation}~(two-excitation QST in $6\times6$ network).

\section{Effects of residual thermal populations}

In both the standard theoretical protocol~\cite{Christandl2004} and our Monte Carlo annealing process, the underlying assumption is that the medium for transferring quantum information is always in its ground state, where no unwanted excitation exists. However, this is not true for real-world superconducting quantum devices. Despite being mounted in a dilution refrigerator with a base temperature of $\sim$20 mK~\cite{yao2023np}, spurious excitations often occur in the qubits~\cite{Jin2015} due to the black-body radiation from higher temperature parts or the thermal heating by control signals. In this section, we investigate the infidelity of quantum state transfer induced by residual thermal populations in the device. 

\begin{figure}[h]
\includegraphics[width=1.0\columnwidth]{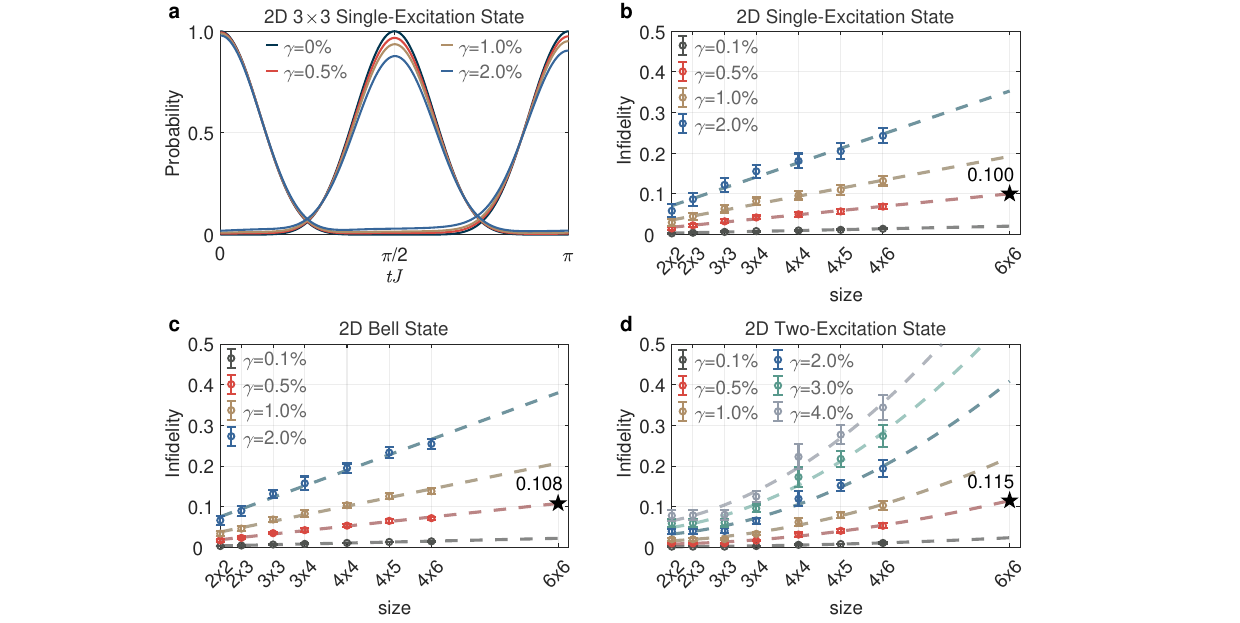}
    \vspace{-0.7cm}
    \caption{\textbf{Effects of thermal populations on QST for different system sizes.}
    \textbf{a}, Simulated population dynamics of  initial qubit $Q_1$ and target qubit  $Q_9$ for the $3\times3$ single-excitation QST, contrasting residual thermal populations of $\gamma=0\%,0.5\%,1\%,$ and $2\%$. The result for each $\gamma$ is the average over 25 random realizations.
    System-size dependence of infidelity computed at $tJ=\pi/2$ for transferring a single-excitation state (\textbf{b}), 
    a Bell state (\textbf{c}), and 
    the two-excitation state (\textbf{d}).
    In the numerical simulations, we use ideal 2D lattices without defects or cross-couplings. We adopt the coupling protocol of Ref.~\cite{Christandl2004} for single-excitation and Bell state QST, and  Monte Carlo annealing protocol for transferring two-excitation state. In panels \textbf{b}, \textbf{c}, and \textbf{d}, circle markers are the numerical results, and the dashed lines are the fitting results. Single-excitation and Bell state results are fitted with a linear function, while two-excitation results are fitted with a quadratic function. Each data point is the average over 25 random realizations, and error bars are the standard deviation. Black stars indicate the extrapolated QST infidelities in the  $6\times6$ network with $\gamma=0.5\%$.}
    \label{subfig:thermalPop_with_size}
\end{figure}

Since a 1D chain with nearest-neighbor $XY$ couplings is always an integrable Hamiltonian irrespective of the population, QST across the chain is insensitive to small thermal populations, as has been reported in Ref.~\cite{Li2018}. The situation is different for the 2D $ XY$ model with cross-couplings that we explore. This system is chaotic in the presence of many excitations and therefore tends to suppress the QST once unwanted thermal populations arise. To quantify such effects on large 2D networks, we numerically simulate small networks of various system sizes and then scale the results to large systems beyond the reach of our current classical computational power. For simplicity,  we model the residual thermal population $\gamma$ by setting the state of each qubit as $|\phi_0^i\rangle = \sqrt{1-\gamma_i}|0\rangle + e^{{\rm i}\theta_i}\sqrt{\gamma_i}|1\rangle$ ($\sqrt{\gamma_i}|0\rangle + e^{{\rm i}\theta_i}\sqrt{1-\gamma_i}|1\rangle$ ) for qubit $Q_i$ expected to be $|0\rangle$ ($|1\rangle$) in initial state. $\gamma_i$ is randomly sampled from a  Gaussian distribution of $N(\gamma,~0.2\gamma)$ and  $\theta_i$ is a random phase uniformly sampled from 0 to $2\pi$. 
The Bell state in qubit pair $(Q_1,Q_2)$ is generated by applying an ideal state-preparation quantum circuit (see Fig.~\ref{subfig:XEB}$\bf a$) to the ground state with thermal populations 
$\bigotimes_{i=1}^2 \left(\sqrt{1-\gamma_i}|0\rangle + e^{{\rm i}\theta_i}\sqrt{\gamma_i}|1\rangle\right)$.
Figure~\ref{subfig:thermalPop_with_size}\textbf{a} shows the population dynamics of single-excitation QST in the 3$\times$3 network for $\gamma\in\{0, 0.5\%, 1.0\%, 2.0\%\}$. As expected, the fidelity of QST is significantly compromised as $\gamma$ increases. 

\begin{figure}[h]
\includegraphics[width=1.0\columnwidth]{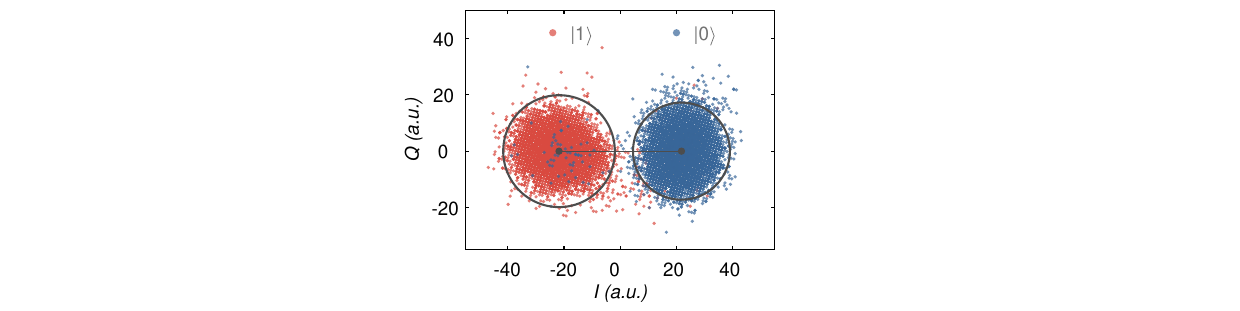}
    \vspace{-0.8cm}
    \caption{\textbf{Measurements of thermal populations.} 
We estimate the residual thermal populations by preparing the qubit in $|0\rangle$ ($|1\rangle$) state and promptly measuring it. Repeating this process 12,000 times for $|0\rangle$ ($|1\rangle$) state, we plot the demodulated readout signals on the I-Q plane~\cite{Jeffrey2014PRL}, where a.u. is the abbreviation of arbitrary units, which means the obtained signal amplitude is not the exact value of the real-world signal but proportional to it. Each blue (red) point represents a single measurement of  $|0\rangle$ ($|1\rangle$) state. About $0.5\%$ of blue points fall into the red region of $|1\rangle$ state, which approximately represents the residual thermal populations. 
    } 
    \label{subfig:thermalPop}
\end{figure}

We further investigate the effects of residual thermal populations for different system sizes. Figures~\ref{subfig:thermalPop_with_size}\textbf {b}, \textbf {c}, and \textbf {d} show the numerical results of the infidelity caused by thermal population $\gamma$ as a function of system size. For all three cases (single-excitation, Bell state, and two-excitation), the infidelity grows as the system size increases. Here, the infidelity is defined  by $(F_{\gamma=0}- F_{\gamma})/F_{\gamma=0}$, where $F_\gamma$ is the transfer fidelity with thermal population of $\gamma$.  To obtain the influence of thermal populations on QST in large systems, such as a  $6\times6$ network, which cannot be simulated efficiently with classical computers,  we perform a system size extrapolation by fitting the relation between the infidelity and the system size using the results of a small number of qubits. As a result, we can approximately estimate the infidelity caused by thermal populations for our experiments, even in the case of 36 qubits. As the typical value of residual thermal population after initialization is $\sim0.5\%$  (see Fig.~\ref{subfig:thermalPop}) for our system, we can deduce that the thermal population induced infidelity of QST in the $6\times6$ network is $\sim10\%$.

\section{Noise analysis for quantum state transfer}

\begin{figure*}[h]
\includegraphics[width=1.0\columnwidth]{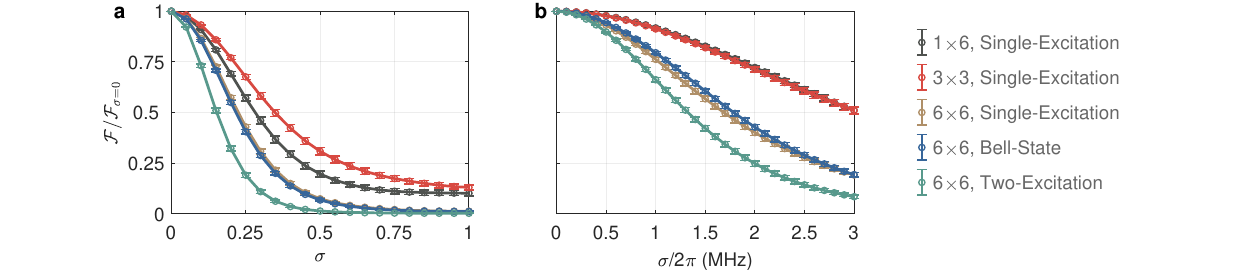}
    \caption{\textbf{Noise analysis for QST.}
    \textbf {a}, Relative QST fidelity compared to the clean case after adding $\{\delta_{ij}J_{ij}\}$ to the NN couplings $\{ J_{ij}\}$ for different cases studied in the main text, where $\delta_{ij}$ is sampled from a Gaussian distribution $N(0, \sigma)$. $6\times 6$ two-excitation case shows the most sensitive trend under the same coupling noise deviation $\sigma$ compared with other cases.
    \textbf {b},  Relative QST fidelity compared to the clean case after adding Gaussian noise $\delta \sim N(0,\sigma)$ to the qubit frequencies. A similar trend is observed, where the $6\times 6$ two-excitation displays the largest decay in comparison to the ideal scenario. Each data point is averaged over 200 random instances, and error bars represent the standard error of the statistical mean.} 
    \label{subfig:noise}
\end{figure*}

On top of the effects of thermal populations, other influences can also affect the experimental realization of an efficient QST, including imperfect experimental control on couplings and qubit frequencies. To account for these, we numerically explore these effects on different cases of QST (see Fig.~\ref{subfig:noise}). We assume a Gaussian-distributed noise $\delta \sim N(0,\sigma)$, where $\sigma$ is the standard deviation. For the noise on the originally optimized NN couplings $\{J_{ij}\}$, we sample $\{\delta_{ij}\}$  from $N(0,\sigma)$ and set the noise-affected NN couplings as \{$(1+\delta_{ij})J_{ij}\}$. By averaging the numerical results over 200 instances, we show that the simulated QST fidelities decay as a function of the standard deviation $\sigma$ in Fig.~\ref{subfig:noise}{\bf a} for all cases considered in the main text. As expected, the noise more easily affects the two-excitation case since it is on the verge of decaying to quantum chaotic behavior. This analysis demonstrates that the solutions for the couplings optimized by the annealing process are rather special.

For the case of noise in the qubit's frequency, the Hamiltonian is modified by adding diagonal terms of the form $\sum \hat\sigma^+_i \hat\sigma^-_i \delta_i$. Here, $\sigma^+_i$ and  $\sigma^-_i$ are raising and lowering operators for the $i$-th qubit. For a given standard deviation $\sigma$, frequency disorders $\{\delta_i\}$ are randomly selected. By averaging over 200 such instances, we obtain the corresponding fidelities of the target qubit(s) (see Fig.~\ref{subfig:noise}{\textbf{b}}). Similar to the noise in the couplings, the QST-fidelity of the two-excitation case is quickly compromised for large $\sigma$.

\section{Measurement of trajectory in large-spin representation}
\begin{figure}[h]
\includegraphics[width=1.0\columnwidth]{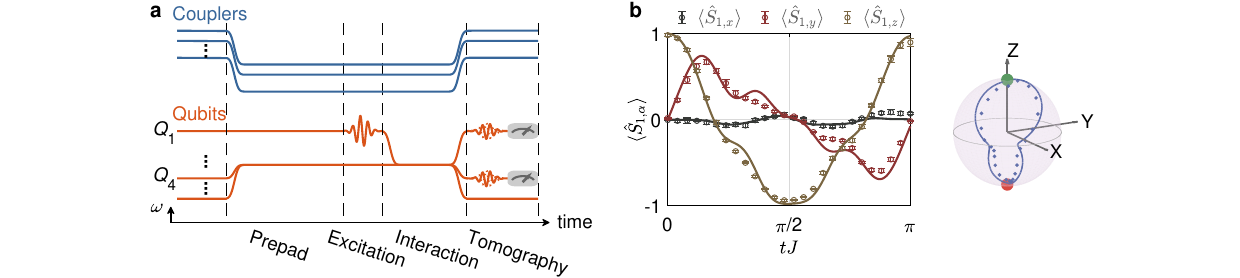}
    \caption{\textbf{Trajectories of QST in large-spin representation.}
    \textbf{a}, Cartoon schematic of typical pulse sequences for two-qubit tomographic measurements of QST dynamics.  Note that we take $Q_{\rm 1}$ and $Q_{\rm 4}$ as an example here. The $\omega$ axis indicates the frequency domain of couplers (blue curves) and qubits (orange curves).  The time axis includes four steps.  First, we apply square pulses on all other qubits except $Q_1$ and all couplers non-neighbor to $Q_1$  to bring them into QST work point. To suppress the small pulse distortion caused by step responses, which may disturb the QST process, we wait for 2~$\mu$s (prepad length) before exciting $Q_1$ with a $\pi$ pulse (orange Gaussian-type waveform) and tuning it to interaction frequency $\omega_I$.  After a QST with a time of $t$,  we apply microwave pulses, selected from $\{I,X/2, Y/2\}^{\otimes 2}$ to target qubits,  $Q_1$ and $Q_4$, for tomographic measurements.
    \textbf{ b}, Measured  trajectories for  $\hat{S}_1$ in the 3$\times$3  network with the QST-optimized couplings (see Fig.~\tcr{2}{\bf g} of the main text).  Left panel shows the dynamics of $\langle\hat{S}_{1,x}\rangle,  \langle\hat{S}_{1,y}\rangle$, and $\langle\hat{S}_{1,z}\rangle$.  The right panel shows the corresponding trajectory on the Bloch sphere.  Circles and solid lines represent experimental and numerical results, respectively. Error bars are the standard deviation of five experimental repetitions.} 
    \label{subfig:tomo phase}
\end{figure}

In the main text, we map single-excitation QST in a 2D network to the precession of two large spins under magnetic fields. We experimentally probe the dynamical trajectories of the two large spins by measuring their projections in the $x,y,z$ directions of the Bloch sphere. Taking single-excitation QST in a $3\times 3$ network as an example~(Figs.~\tcr{2}\textbf{f} and \textbf{i} of the main text), the matrix representations of  the spin operators $\hat{S}_{1,\alpha}$ and $\hat{S}_{2,\alpha}$ $(\alpha=x,y,z)$ are
\begin{align}
&\hat{S}_{1,x}=\frac{1}{\sqrt{2}}\left (\begin{array}{rrrr}
0 &1   & 0 \\
1 &0& 1  \\
0 & 1&0 \\
\end{array}\right)\otimes 
\left (\begin{array}{rrrr}
1&0   & 0 \\
0 &1& 0  \\
0 &  0&1 \\
\end{array}\right) \ , \ \
\hat{S}_{1,y}=\frac{1}{\sqrt{2}}\left (\begin{array}{rrrr}
0 &-i   & 0 \\
i &0& -i  \\
0 & i &0 \\
\end{array}\right)\otimes 
\left (\begin{array}{rrrr}
1&0   & 0 \\
0 &1& 0  \\
0 &  0&1 \\
\end{array}\right)  \ , \ \
\hat{S}_{1,z}=\left (\begin{array}{rrrr}
1 &0   & 0 \\
0 &0& 0  \\
0 & 0 &-1 \\
\end{array}\right)\otimes 
\left (\begin{array}{rrrr}
1&0   & 0 \\
0 &1& 0  \\
0 &  0&1 \\
\end{array}\right) \ , \\
&\hat{S}_{2,x}=\left (\begin{array}{rrrr}
1&0   & 0 \\
0 &1& 0  \\
0 &  0&1 \\
\end{array}\right)
\otimes \frac{1}{\sqrt{2}}\left (\begin{array}{rrrr}
0 &1   & 0 \\
1 &0& 1  \\
0 & 1 &0 \\
\end{array}\right) \ , \ \
\hat{S}_{2,y}=\left (\begin{array}{rrrr}
1&0   & 0 \\
0 &1& 0  \\
0 &  0&1 \\
\end{array}\right)
\otimes \frac{1}{\sqrt{2}} \left (\begin{array}{rrrr}
0 &-i   & 0 \\
i &0& -i  \\
0 & i &0 \\
\end{array}\right) \ , \ \
\hat{S}_{2,z}=\left (\begin{array}{rrrr}
1&0   & 0 \\
0 &1& 0  \\
0 &  0&1 \\
\end{array}\right)\otimes \left (\begin{array}{rrrr}
1 &0   & 0 \\
0 &0& 0  \\
0 & 0 &-1 \\
\end{array}\right) \ .
\end{align}

A quantum state  $|{\psi}\rangle$ of the two spins has the following expectation values
\begin{align}
\langle\hat{S}_{1,x}\rangle \ , \ \  \langle\hat{S}_{1,y}\rangle \ , \ \ \langle\hat{S}_{1,z}\rangle \ , \ \
\langle\hat{S}_{2,x}\rangle \ , \ \ \langle\hat{S}_{2,y}\rangle \ , \ \
\langle\hat{S}_{2,z}\rangle \ ,
\end{align}
where the expectation value $\langle\hat{O}\rangle$ of an operator $\hat{O}$ is given by $\langle O\rangle=\langle\psi|\hat{O}|\psi\rangle$.  In the single-excitation-conserved subspace of 9 qubits,  $|{\psi}(t)\rangle=\sum_{i=1}^9 a_i(t) |\phi_i\rangle$, where $|\phi_i\rangle=(0,\ldots ,1,\ldots ,0)^{\rm T}$ is a basis vector of the subspace, such that the $i$-th qubit is excited while the remaining ones are in their ground states. We can thus conveniently represent expectation values of the spin operators using the coefficients of $|\psi\rangle$.  Taking $\hat{S}_{1,x}$ 
for example, we have 
\begin{align}
\begin{split}
\langle\hat{S}_{1,x}\rangle=&
    (a_1{a_4}^*+a_4{a_1}^*+a_2{a_5}^*+a_5{a_2}^*\\ 
    &+a_3{a_6}^*+a_6{a_3}^*+a_4{a_7}^*+a_7{a_4}^*\\ 
    &+a_5{a_8}^*+a_8{a_5}^*+a_6{a_9}^*+a_9{a_6}^*)/\sqrt{2}.
\end{split}
\end{align}
Each term  $a_i{a_j}^*$  corresponds to an element of a two-qubit density matrix $\rho_{(Q_i, Q_j)}$, which can be easily extracted using tomographic measurements.

Figure~\ref{subfig:tomo phase}{\bf a} shows an example of experimental sequences for the tomographic measurements of the qubit pair $(Q_1, Q_4)$.  The measurements for other qubit pairs are similarly obtained by applying the tomographic pulses to the corresponding qubits. Figure~\ref{subfig:tomo phase}{\bf b}  presents the experimentally measured dynamics of  $\langle\hat{S}_{1,x}\rangle, \langle\hat{S}_{1,y}\rangle, \langle\hat{S}_{1,z}\rangle$ and the trajectories on the Bloch sphere for the QST in $3\times3$ network with the optimized coupling configurations (see Figs.~\tcr{3}\textbf{h} and \textbf{i} of the main text), which are in good agreement with the numerics. 

\section{Preparation of Bell state}

\begin{figure}[h]
\includegraphics[width=1.0\columnwidth]{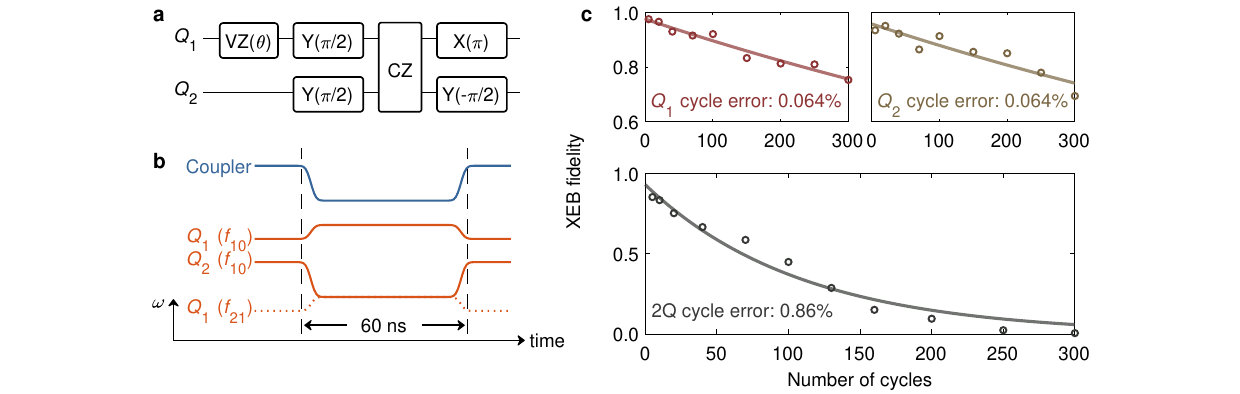}
    \caption{\textbf{Preparation of the Bell state with CZ gate.}
    \textbf {a}, Quantum circuits for preparing a Bell state between $Q_{\rm 1}$ and $Q_{\rm 2}$. $Y(\theta)$~($X(\theta)$) denotes rotating the state an angle $\theta$ around the $y$-axis ($x$-axis) of the Bloch sphere.
    \textbf{b}, Pulse sequences for implementing a  two-qubit CZ gate assisted with a tunable coupler.  The gate time is about 60 ns in the experiment.
    \textbf{c}, Cross-entropy benchmarking of single-qubit gates and two-qubit CZ gate. Pauli errors of single-qubit gates for  $Q_{\rm 1}$ and  $Q_{\rm 2}$ are benchmarked by simultaneous XEB. For the CZ gate, each cycle contains two parallel single-qubit gates and a subsequent CZ gate. Pauli error of CZ gate extracted here is $\sim0.73\%$.
    } 
    \label{subfig:XEB}
\end{figure}

The initial Bell state in Fig.~\tcr{3}\textbf{b} of the main text is prepared by applying a quantum circuit to  $Q_{\rm 1}$ and $Q_{\rm 2}$, which is shown in Fig.~\ref{subfig:XEB}\textbf{a}. The entanglement is generated by a two-qubit control-Z~(CZ) gate,  which is implemented by dynamically tuning on/off the interaction between qubit energy levels of $|11\rangle$ and $|20\rangle$. The corresponding pulse sequences are shown in Fig.~\ref{subfig:XEB}\textbf{b}. A good preparation of a Bell state largely relies on the quality of CZ and single-qubit gates. To characterize them, we use the cross-entropy benchmarking (XEB) technique~\cite{Arute2019}. The results are shown in Fig.~\ref{subfig:XEB}\textbf{c}. Pauli errors of $\sim$0.064\%  and $\sim0.73\%$ are extracted for single-qubit gates and CZ gate, respectively. The prepared initial Bell state is further confirmed by the reconstructed density matrix using quantum state tomography~(see Fig.~\tcr{3}\textbf{b} of main text), which yields a state fidelity of  $\sim$ 0.992. 

Due to the finite step edge times of square pulses for realizing QST, the Bell state can accumulate an unwanted dynamical phase during the frequency shifts of $Q_1$ and $Q_2$. To guarantee the initial Bell state for QST is $|\Psi^{-}\rangle = (|01\rangle - |10\rangle)/{\sqrt{2}}$ at the beginning of interaction, a virtual Z  phase gate (VZ($\theta$)) \cite{McKay2017VZgate} is applied to $Q_{\rm 1}$ to compensate it.

\section{(Non)Ergodicity of two-excitation Hamiltonians}

\begin{figure}
\includegraphics[width=1.0\columnwidth]{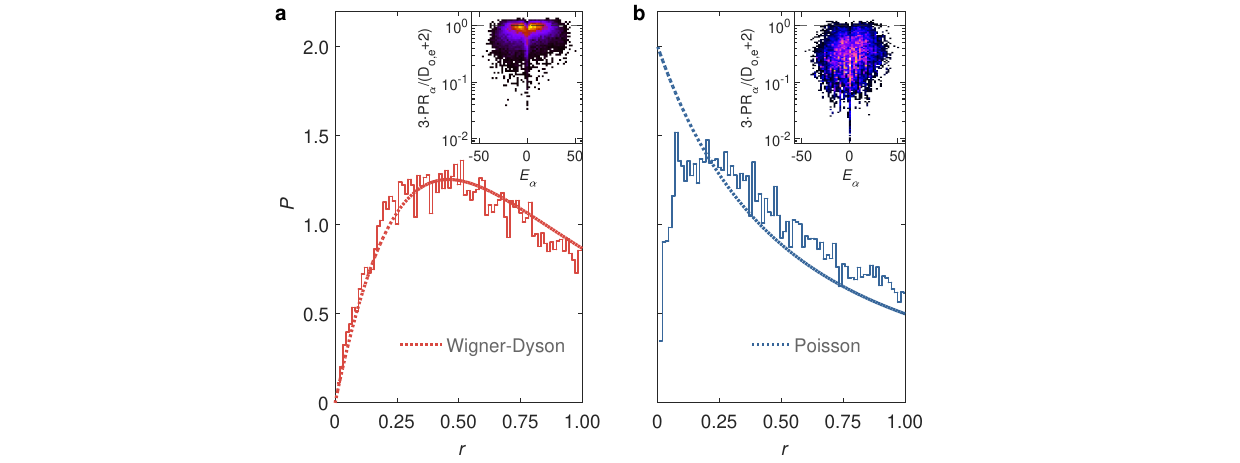}
    \caption{
    {\bf Ergodicity analysis of random and optimized couplings for QST.} 
    The distribution of the ratio of adjacent gaps $r$ for two excitations in the $6\times 6$ qubit network:  {\bf a}, for the case of random couplings, and  {\bf b}, for the solutions extracted from the optimization process that maximizes the quantum state transfer at later times. Owing to the inversion symmetry of the couplings (nearest-neighbors or across plaquettes), even (e) and odd (o) sectors are resolved. Dashed lines give the Wigner surmise for $P(r)$ and the Poisson distribution, respectively. Forty realizations of both random couplings and optimized solutions are averaged. The corresponding insets give a two-dimensional histogram of the participation ratio of each eigenstate $|\alpha\rangle$ in the computational basis for the whole ensemble of solutions, normalized by the Hilbert dimensions ${\cal D}_{o, e}$ [${\cal D}_{o} = 306$ and ${\cal D}_{e} = 324$]; brighter colors are associated with higher counts, and horizontal lines mark the GOE prediction extracted from the theory of random matrices. The matrices used in both cases have bounds on the nearest-neighbor couplings given by  $[J_{\rm min}, J_{\rm max}] = [-10, -0.1]$ MHz.
    }
    \label{subfig:QST_erg}
\end{figure}

An often-used analysis to classify quantum chaotic properties of Hamiltonians of interest is to investigate the degree of level repulsion of their eigenvalues $\{E_\alpha\}$. While typical ergodic Hamiltonians exhibit a Wigner-Dyson-like distribution $P(s)$ for the gaps between nearest eigenvalues ($s_\alpha \equiv E_{\alpha+1}-E_\alpha$), non-ergodic ones show, on the other hand, absence of level repulsion; the statistical properties encoded in $P(s)$ follow a Poisson distribution instead~\cite{Porter1962}. To avoid the complication of unfolding the spectrum to guarantee a unit mean-level spacing, we take instead the ratio of adjacent gaps $r_\alpha \equiv \min(s_\alpha,s_{\alpha+1})/\max(s_\alpha,s_{\alpha+1})$~\cite{Oganesyan2007}. In this case, the ergodic and non-ergodic (Poisson) distributions translate, respectively, to~\cite{Atas2013}
\begin{align}
    P_{\rm GOE}(r) = \frac{27}{8}\frac{r + r^2}{(1+r+r^2)^{5/2}} \ , \ \ \ \  P_{\rm P}(r) = \frac{2}{1+r^2}\ ,
\end{align}
where we focus on the case of the class of random matrices belonging to the Gaussian orthogonal ensemble (GOE), which are real and symmetric.

A complication in this exploration arises when existing symmetries of the Hamiltonian spoil the classification of its ergodic properties if they are not resolved~\cite{Mondaini2018}. In particular, as explained in the main text, we enforce a real-space inversion-symmetry constraint for the Hamiltonians we engineer, which, in practice, makes its spectrum independently subdivided into sectors whose parity is even or odd under this point group symmetry. By independently constructing the Hamiltonian in each of these subsectors, we contrast the gap distributions for the cases where the couplings are either randomly selected within a uniform distribution $[J_{\rm min}, J_{\rm max}]$ or when they are optimized (within the same bounds) to achieve a quantum state transfer at times $t_{\rm QST} = \pi/(2 J)$; the corresponding distributions $P(r)$ are reported in Fig.~\ref{subfig:QST_erg}. They show that for solutions that optimize the QST, here selected to have fidelities ${\cal F}>0.93$, $P(r)$ approaches a Poisson distribution (i.e., quasi-nonergodic), unlike the cases with random couplings, more closely similar to the GOE surmise. 

Additionally, we investigate ergodic properties of the eigenstates themselves (insets in Fig.~\ref{subfig:QST_erg}), via the quantification of the participation ratio in the computational basis (inversion-symmetric Fock states $|n\rangle$)
\begin{align}
    {\rm PR}_\alpha = \frac{1}{\sum_{n = 1}^{{\cal D}_{o, e}} |c_\alpha^n|^4}\ ,
\end{align}
where $c_\alpha^n = \langle n|\alpha\rangle$. Random matrices of the GOE class exhibit ${\rm PR} = \frac{{\cal D}+2}{3}$~\cite{Izrailev1990, Zelevinsky1996}. The typical eigenstates from the Hamiltonian with optimized couplings deviate from this prediction much more markedly than do those with random couplings, as seen in the insets of Fig.~\ref{subfig:QST_erg}, showing a two-dimensional histogram of all eigenstates for the forty realizations considered and noting the logarithmic scale.

This analysis makes it immediately apparent that when random couplings are used, the ensuing spectral properties exhibit characteristics of ergodic systems; the same would be the case even if homogeneous couplings in our Hamiltonian were used instead (if all point-group symmetries were resolved). Conversely, optimizing the coupling parameters to achieve a high-fidelity QST converges to coupling matrices $\{J_{ij}\}$ whose associated Hamiltonian avoids this ergodic fate and steers it towards nonergodicity.

A physical picture is useful to understand this difference. In ergodic Hamiltonians, one expects that the evolution governed by $e^{-{\rm i}\hat H t}$ leads to a diffusive exploration of the Hilbert space with time $t$ when the initial state is a single point in this space (Fock state). In contrast, non-ergodic Hamiltonians typically exhibit characteristic revivals throughout the dynamics, leading them to the possibility of exploring states arbitrarily close to the initial conditions, for example. Similarly, one can engineer (quasi) non-ergodic Hamiltonians such that at a certain time, the majority of the contribution to $|\psi(t)\rangle$ resides in another single Fock state (point in the Hilbert space). This is the procedure envisaged by the quantum state transfer.

\begin{figure*}[h]
\includegraphics[width=0.9\columnwidth]{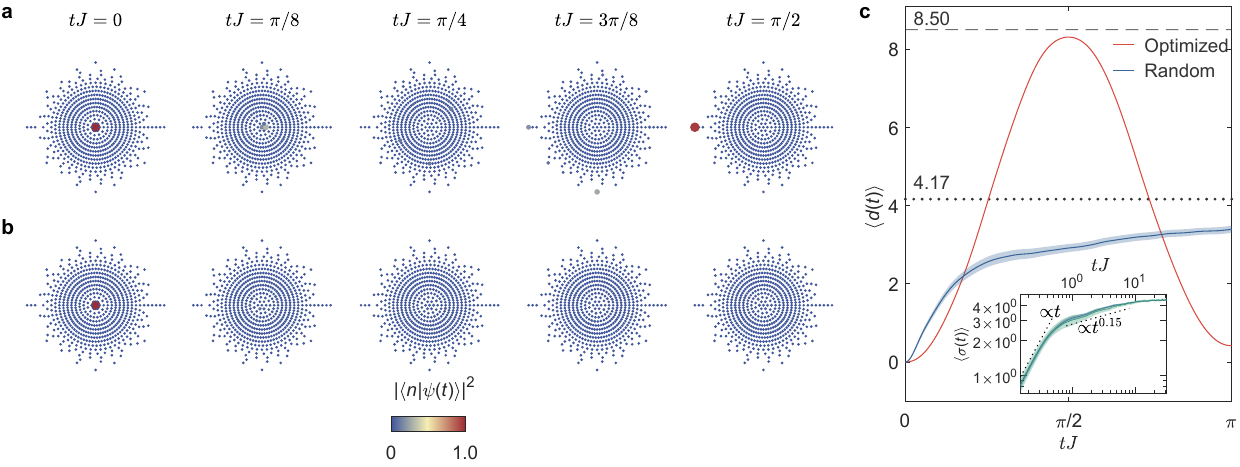}
    \caption{{\bf Dynamics of wave-packets in Fock space.}
    {\bf a,} Projection of the time-evolved wavefunction in Fock space, $|\langle n| \psi(t)\rangle|^2$ for the case where the couplings $\{J_{i,j}\}$ were optimized to maximize the quantum state transfer at times $t J = \pi/2$ [$J=1$ MHz here], taking into account both the cross-couplings $J^{\times}$ and the defective coupler; different time-snapshots are marked. Here, the radial direction is given by $d(|0\rangle, |n\rangle)$ (see text) while the angular one is arbitrary; $|\langle n| \psi(t)\rangle|^2$ is mapped by the color bar and the size of the marker at each Fock state $|n\rangle$. {\bf b,} The same for the case of random couplings: Here, the wave-packet diffuses in Fock space. {\bf c,} The dynamics of the wave-packet's `center of mass' for optimized and random couplings; the inset gives the wave-packet spread in time for the case of the random couplings, contrasting $6\times 6$ (blue) and $12\times 12$ (green), (averaged over 40 and 10 realizations, respectively), with regimes marking an initial ballistic transient and another of subdiffusive spreading. The horizontal dashed (dotted) line in the main panel depicts the $d(|0\rangle,|n_{\rm target}\rangle)$ ($\overline d$) value -- see text. The matrices used in both cases have bounds on the nearest-neighbor couplings given by  $[J_{\rm min}, J_{\rm max}] = [-10, -0.1]$ MHz.}
    \label{subfig:Mhd}
\end{figure*}

To illustrate this, we show in Figs.~\ref{subfig:Mhd}{\bf a,} and \ref{subfig:Mhd}{\bf b} the dynamics of an initially prepared two-excitation state with optimized and random couplings, respectively. This is schematically seen as the weight of $|\psi(t)\rangle$ in each Fock state $|n\rangle$ for various times. This expands the original description in the main text, but now for a $6\times 6$ qubit network, and its corresponding ${\cal D}_{\hat H} = \binom{36}{2} = 630$ Fock states. While the QST is seen in the case of optimized couplings, manifested as a revival of the maximum weight in a Fock state at times $tJ = \pi/2$, random couplings fail to have an instant of time where the projection is accumulated in any single $|n\rangle$ at $t>0$. It indicates that the dynamics occur diffusively over the Fock space if random couplings are taken.

To explore this analogy further, we use a metric for distances between Fock states, introduced in the Methods Section in the main text. For completeness, we partially repeat this description here. The idea is to establish an associated time for a Fock state to be reached, given the typical coupling's matrix values -- a distant Fock state would thus take numerous hopping times to be accessed. If using the initial state as a reference, we defined a distance as $d(|0\rangle,|n^\prime\rangle) = \frac{1}{4}\sum_{l=1}^2\left(|x_l^\prime - x_0| + |y_l^\prime - y_0| + |x_l^\prime - x_1| + |y_l^\prime - y_1|\right) - 1/2$, where $(x_l^\prime,y_l^\prime)$ are the cartesian coordinates of each of the $l$-excitations ($l=2$) of a generic Fock state $|n^\prime\rangle$. For the initial state $|n=0\rangle$, one thus have $(x_0,y_0)$ and $(x_1,y_1)$ being the coordinates of its excitations. Hence $d(|0\rangle,|0\rangle) = 0$ whereas $d(|0\rangle,|n_{\rm target}\rangle) = 8.5$ for the target state in this network size.

An average distance can be dynamically defined as
\begin{equation}
    \langle d(t)\rangle = \sum_{n=0}^{630-1} d(|0\rangle, |n\rangle)\  |\langle n|\psi(t)\rangle|^2\ ;
\end{equation}
likewise, the root-mean-square $\sigma$ of the wave-packet spreading in the Fock basis is 
\begin{equation}
    \langle \sigma(t)\rangle = \sqrt{\sum_{n=0}^{630-1} d^2(|0\rangle, |n\rangle)\  |\langle n|\psi(t)\rangle|^2}\ .
\end{equation}

Figure~\ref{subfig:Mhd}{\bf c} displays the time-evolution of both quantities, contrasting the case of an optimized solution of the couplings with the averaged value over 40 realizations of random couplings instances. The average distance $\langle d(t)\rangle$ measures the evolution of the wave-packet `center of mass' in Fock space: For QST-optimized couplings, one observes a ballistic (almost) periodic evolution between the initial state $|0\rangle$ and the target state $|n_{\rm target}\rangle$. Conversely, a slow evolution towards the mean distance $\overline{d} =\frac{1}{630}\sum_{n=0}^{630-1} d(|0\rangle, |n\rangle)$ is seen in the case of randomly chosen $\{J_{ij}\}$. These results are similar to the theory/experimental ones described in Fig.~\tcr{4} in the main text for a smaller qubit network size, $3\times 3$.

A refined characterization of the propagation in Fock space is given by $\sigma$, which typically exhibits a time-dependence $\propto t^\alpha$. For $\alpha=1/2$($\alpha<1/2$), the wave-packet propagates diffusively (sub-diffusively) in Fock space, while $\alpha = 1$ describes ballistic transport. We note that at short-time scales (inset in Fig.~\ref{subfig:Mhd}{\bf c}), a transient ballistic propagation is observed that gives way to a subdiffusive one at later times before the effects of the subspace's finiteness set in at $tJ \gg 1$. Such a mix of ballistic and diffusive behavior is similarly observed in (excitation) number-conserving random unitary circuits~\cite{Khemani2018}, with random gates. These are not equivalent to the unitary evolution with random couplings that we perform here, that is, with a fixed functional form of the gates but random amplitudes. However, they do share similarities (at least in short times) if one performs a Trotterization of the dynamical evolution. Whether the subdiffusive spread in intermediate time scales we observe (as opposed to diffusive) is affected by the system's finiteness or the small number of excitations we consider deserves future investigation. Likewise, random amplitude unitary circuits being dissimilar from random unitary circuits, the corresponding expected operator spreading, as investigated in Ref.~\cite{Khemani2018}, is hitherto unknown.

\section{Numerically optimized quantum state transfer with Monte Carlo annealing}

\begin{figure}[h]
\includegraphics[width=1.0\columnwidth]{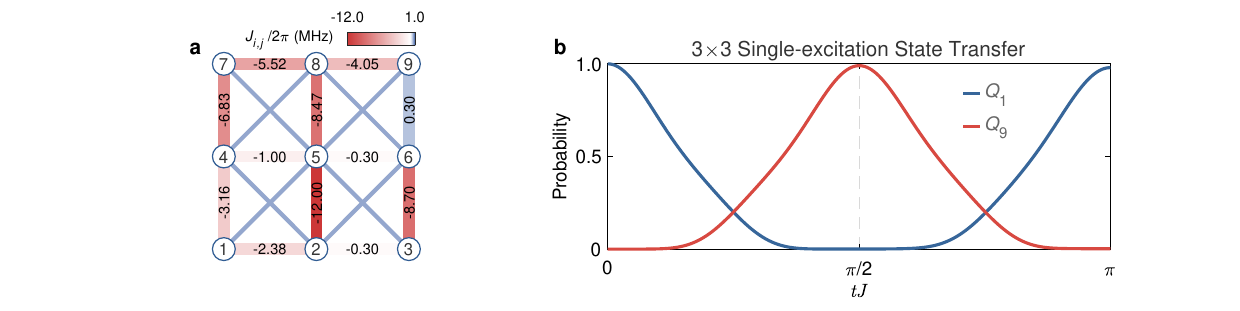}
\vspace{-0.8cm}
    \caption{\textbf{Numerically optimized couplings for single-excitation QST in the $3\times3$ network.} 
    \textbf {a}, Numerically optimized NN couplings. 
    \textbf {b}, The corresponding dynamics for transferring single excitation from $Q_1$ to $Q_{9}$, with a maximum QST fidelity of $\sim$ 0.9902. 
    }
    \label{subfig:ideal_3x3_1e_opt}
\end{figure}

\begin{figure}[H]
\includegraphics[width=1.0\columnwidth]{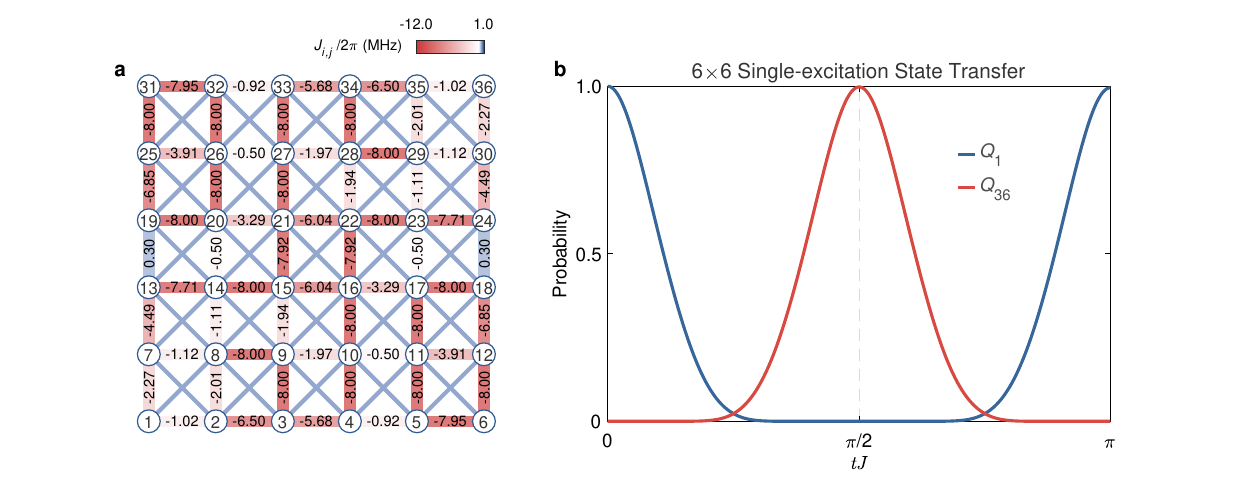}
\vspace{-0.8cm}
    \caption{\textbf{Numerically optimized couplings for single-excitation QST in the $6\times6$ network.} 
    \textbf {a}, Numerically optimized NN couplings. 
    \textbf {b}, The corresponding dynamics for transferring single excitation from $Q_1$ to $Q_{36}$, with a maximum QST fidelity of $\sim$ 0.9979. 
    }
    \label{subfig:ideal_6x6_1e}
\end{figure}

\begin{figure}[H]
\includegraphics[width=1.0\columnwidth]{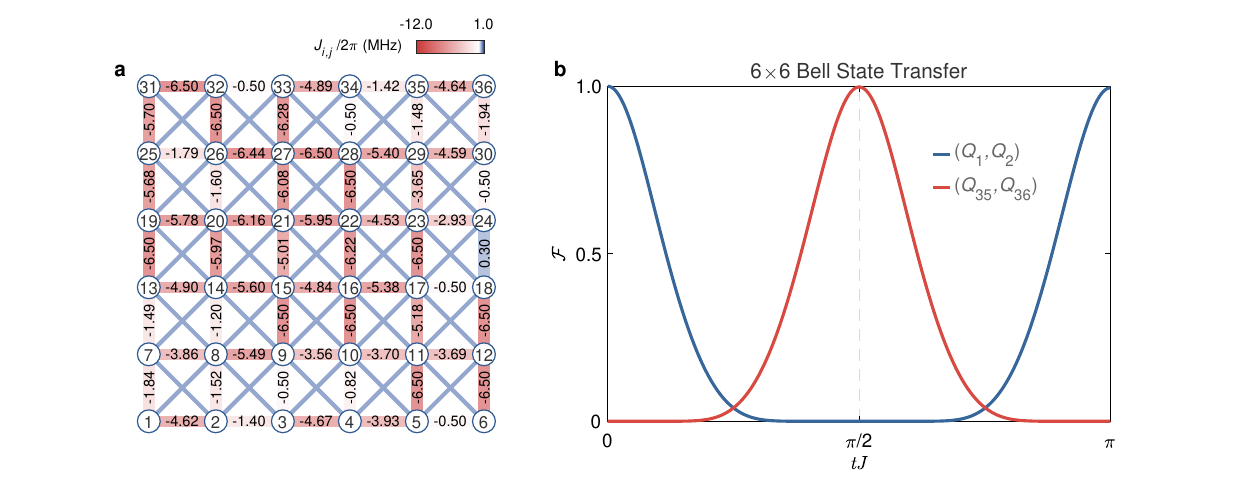}
\vspace{-0.8cm}
    \caption{\textbf{Numerically optimized couplings for Bell state QST in the $6\times6$ network.} 
    \textbf {a}, Numerically optimized NN couplings. 
    \textbf {b}, The corresponding dynamics for transferring Bell state from $(Q_1,Q_2)$ to $(Q_{35},Q_{36})$, with a maximum QST fidelity of $\sim$ 0.9978. 
    }
    \label{subfig:ideal_6x6_bell}
\end{figure}

\begin{figure}[H]
\includegraphics[width=1.0\columnwidth]{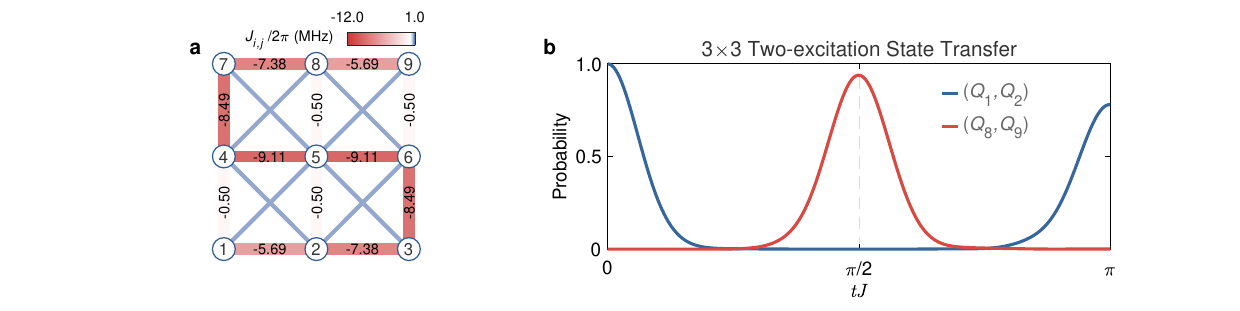}
\vspace{-0.8cm}
    \caption{\textbf{Numerically optimized couplings for two-excitation QST in the $3\times3$ network.} 
    \textbf {a}, Numerically optimized NN couplings. 
    \textbf {b}, The corresponding dynamics for transferring two excitations from $(Q_1,Q_2)$ to $(Q_{8},Q_{9})$, with a maximum QST fidelity of $\sim$ 0.9388. 
    }
    \label{subfig:ideal_3x3_2e_opt}
\end{figure}

\begin{figure}[H]
\includegraphics[width=1.0\columnwidth]{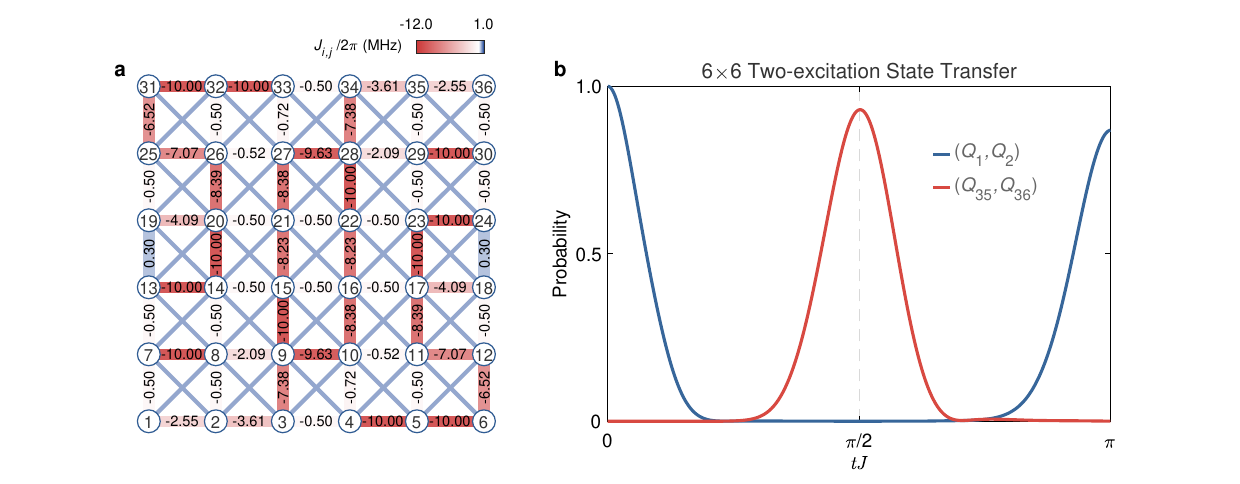}
\vspace{-0.8cm}
    \caption{\textbf{Numerically optimized couplings for two-excitation QST in the $6\times6$ network.} 
    \textbf {a}, Numerically optimized NN couplings.
    \textbf {b} The corresponding dynamics for transferring two excitations from $(Q_1,Q_2)$ to $(Q_{35},Q_{36})$, with a maximum QST fidelity of $\sim$ 0.9301. 
    }
    \label{subfig:ideal_6x6_2e}
\end{figure}

In the main text, to realize a high-fidelity QST in a large, imperfect two-dimensional quantum network where unwanted cross-couplings and a defective coupler exist, we employ a Monte Carlo annealing procedure to optimize NN couplings of the network. With the numerically optimized result as a reference, we experimentally calibrate our superconducting quantum processor to demonstrate the quantum state transfer of few excitations and the physical insight behind it. In this section, we provide the optimized coupling values and the corresponding QST dynamics for the cases of single-excitation (see Fig.~\ref{subfig:ideal_3x3_1e_opt} and Fig.~\ref{subfig:ideal_6x6_1e}), Bell state (see Fig.~\ref{subfig:ideal_6x6_bell}), and two excitations (see  Fig.~\ref{subfig:ideal_3x3_2e_opt} and Fig.~\ref{subfig:ideal_6x6_2e}). In two of these solutions, we release the inversion symmetry constraint to achieve a slightly higher QST-fidelity, given that more degrees of freedom is beneficial for the optimization procedure.

\section{Quantum speed limit bounds}

In the main text, we argue that the quantum state transfer cannot occur arbitrarily fast but rather needs to, at least, take place at times equal to the minimum orthogonalization time, whose values are shown in the Figs. \tcr{2}, \tcr{3}, and \tcr{4} of the main text. Such an orthogonalization time may depend on either the (time-conserved) mean energy $E$ (with respect to the ground-state energy) or the energy uncertainty $\Delta E$. In addition, the actual overlap of the time-evolving wave function with the initial state is bounded, as demonstrated by Margolus and Levitin (ML)~\cite{Margolous1998} for the case of the mean energy-bound,
\begin{equation}
    |\langle \psi(0) | \psi(t)\rangle| \geq \cos\left(\sqrt{\frac{\pi E t}{2\hbar}}\right),
    \label{eq:ML_bound}
\end{equation}
or in the $\Delta E$-bounded dynamics, as shown by Mandelstam and Tamm (MT)~\cite{Mandelstam1944}
\begin{equation}
    |\langle \psi(0) | \psi(t)\rangle| \geq \cos\left(\frac{\Delta E t}{\hbar}\right)\ .
    \label{eq:MT_bound}
\end{equation}

\begin{figure}[H]
\includegraphics[width=1.0\columnwidth]{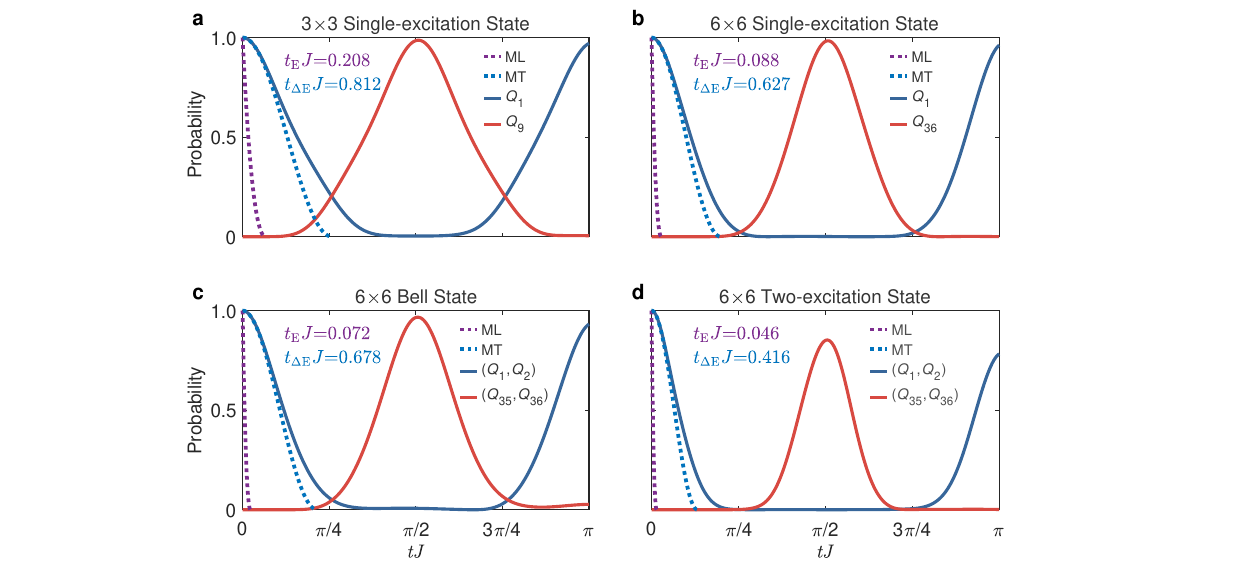}
    \caption{
    {\bf Quantum speed limit bounds for 2D optimized couplings.} 
    {\bf a-d}, Analysis of the quantum speed limits bounds [Eqs.~\eqref{eq:ML_bound} and \eqref{eq:MT_bound}] for the QST optimized couplings of {\bf a,} 2D $3\times3$ single-excitation state, {\bf b,} 2D $6\times6$ single-excitation state, {\bf c,} 2D $6\times6$ Bell state and {\bf d,} 2D $6\times6$ two-excitation state. The actual population dynamics for the initial and final qubits are also included. The former is limited from below by the MT bound. The values of $t_{\Delta E}$ and $t_E$ annotated in each panel stand for the minimal time given by the quantum speed limit of the corresponding MT and ML bounds, respectively.
    } 
    \label{subfig:QSL_bound}
\end{figure}

To explore this further, we report in Fig.~\ref{subfig:QSL_bound} the dynamics of the four cases studied to accomplish QST. In all situations, the MT-bound governs the dynamics, and the tighter bound is in the case of the two-excitation state. This is seen for various solutions of the optimized couplings that maximize the QST fidelity. Nevertheless, it is important to emphasize that generically, it is always possible to make the dynamics governed by the ML bound, provided that the initial state has associated mean energy sufficiently close to the ground state of the emulated Hamiltonian. For the initial states we considered here, that was never the case.

\bibliography{QST.bib}